\documentclass[12pt,fleqn,a4paper]{article}
\usepackage{graphicx}
\usepackage{latexsym}
\usepackage[small,bf]{caption2}
\usepackage{amsfonts}
\usepackage{amssymb}
\usepackage{amsmath}
\usepackage{bbm}
\usepackage{times}
\usepackage[pdftex]{hyperref}

\usepackage{theorem}
\newtheorem{theorem}{Theorem}
\newtheorem{lemma}{Lemma}
\newtheorem{proposition}{Proposition}
\newtheorem{corollary}{Corollary}

\newcommand{\be}{\begin{equation}}
\newcommand{\ee}{\end{equation}}
\newcommand{\bel}[1]{\begin{equation}\label{#1}}
\newcommand{\bea}{\begin{eqnarray}}
\newcommand{\eea}{\end{eqnarray}}
\newcommand{\bef}{\begin{figure}}
\newcommand{\enf}{\end{figure}}
\newcommand{\ba}{\begin{array}}
\newcommand{\ball}{\begin{array}{ll}}
\newcommand{\bacl}{\begin{array}{cl}}
\newcommand{\bacll}{\begin{array}{cll}}
\newcommand{\bal}{\begin{array}{l}}
\newcommand{\bac}{\begin{array}{c}}
\newcommand{\ea}{\end{array}}
\newcommand{\N}{{\mathbb{N}}}
\newcommand{\R}{{\mathbb{R}}}
\newcommand{\1}{{\mathbbm{1}}}
\newcommand{\feta}{{\boldsymbol{\eta}}}

\begin{document}

\title{Discontinuous condensation transition\\ and nonequivalence of ensembles\\ in a zero-range process}

\author{Stefan Grosskinsky\footnote{Mathematics Institute, Zeeman Building, University of Warwick, Coventry CV4 7AL, UK, S.W.Grosskinsky@warwick.ac.uk} and Gunter M.~Sch\"utz\footnote{Forschungszentrum J\"ulich GmbH, Institut f\"ur Festk\"orperforschung, D-52425 J\"ulich, Germany, G.Schuetz@fz-juelich.de}}

%\author{Stefan Grosskinsky\and Gunter M.~Sch\"utz}

%\institute{Stefan Grosskinsky\at Mathematics Institute, Zeeman Building, University of Warwick, Coventry CV4 7AL, UK, \email{S.W.Grosskinsky@warwick.ac.uk}, phone: +44 2476 522673, fax: +44 2476 524182\and Gunter M.~Sch\"utz\at  Forschungszentrum J\"ulich GmbH, Institut f\"ur Festk\"orperforschung, D-52425 J\"ulich, Germany, \email{G.Schuetz@fz-juelich.de}, phone: +49 2461 616264, fax: +49 2461 612850}

%\begin{document}

\maketitle

\begin{abstract}
We study a zero-range process where the jump rates do not only depend on the local particle configuration, but also on the size of the system. Rigorous results on the equivalence of ensembles are presented, characterizing the occurrence of a condensation transition. In contrast to previous results, the phase transition is discontinuous and the system exhibits ergodicity breaking and metastable phases. This leads to a richer phase diagram, including nonequivalence of ensembles in certain phase regions.
The paper is motivated by results from granular clustering, where these features have been observed experimentally.
\end{abstract}

%\keywords{zero range process\and discontinuous phase transition\and equivalence of ensembles\and metastability\and ergodicity breaking\and granular clustering}
\textbf{keywords.} zero range process; discontinuous phase transition; equivalence of ensembles; metastability; ergodicity breaking; granular clustering

\section{Introduction}

The zero-range processes is an interacting particle system introduced in \cite{spitzer70}, which has recently attracted attention due to the possibility of a condensation transition. A prototype model with space homogeneous jump rates that exhibits condensation has been introduced in \cite{evans00}. When the particle density $\rho$ in the system exceeds a critical value $\rho_c$, the system phase separates in the thermodynamic limit into a homogeneous background with density $\rho_c$ and a condensate, that contains all the excess particles. This phase transition is by now well understood on a mathematically rigorous level for general zero-range processes \cite{stefan}, and has been applied to model clustering phenomena in various fields (see \cite{evansetal05} and references therein). In one dimension, a mapping to exclusion models gives rise to a criterion for 
non-equilibrium phase separation \cite{kafrietal02}. Further rigorous results on the zero-range process include a proof of condensation even on finite lattices \cite{ferrarietal07}, and a refinement of the results in \cite{stefan}, which implies a limit theorem for typical density profiles in case of condensation \cite{loulakisetal07}. Regarding the background density as order parameter, it has been shown in a general context (including different particle species) that in spatially homogeneous zero-range processes with a stationary product measure condensation is always a continuous phase transition \cite{stefan2}. Recently, investigations have been further extended to open boundaries where particles are injected and extracted \cite{levineetal05} and heuristically to various generalized models. Those include a non-conserving zero-range process that exhibits generic critical phases \cite{angeletal07}, zero-range processes with non-monotonic jump rates leading to multiple condensate sites \cite{schwarzkopfetal08}, or mass transport models with pair-factorised stationary measures that give rise to a spatially extended condensate \cite{evansetal06}.

In this paper we study the condensation transition in a generalized zero-range process where the jump rates depend on the system size. The motivation for this study comes from experiments on granular media reported in \cite{schlichtingetal96,weeleetal01,meeretal02}. Granular particles are distributed uniformly in a container which is divided in several compartments. When shaking the container, the particles start clustering in some of the compartments and after equilibration, almost all particles form a "condensate" in one of the compartments. The phenomenon is robust for a variety of shaking strengths and a gas-kinetic approach lead to a simplified model equivalent to a zero-range process where the hopping rates depend on the number of compartments \cite{eggers99,weeleetal01,meeretal02,meeretal07}. In an alternative activated-process approach it can be modeled by a zero-range type process, where the jump rates depend on the total number of particles in the system \cite{lipowskietal02,coppexetal02}, and both approaches have been summarized in \cite{toeroek05}. A heuristic analysis of the behaviour of the order parameter agrees with experimental observations and shows that generically the transition is discontinuous and the system exhibits hysteresis and metastability. This analysis suggests that the discontinuity is due to the dependence of the jump rates on the total number of particles or the number of compartments, respectively.

To treat this phase transition on a rigorous level, we present a detailed analysis of a simple prototype model with system-size dependent jump rates, for which we derive results in the context of the equivalence of ensembles analogous to \cite{stefan,stefan2}. From a mathematical viewpoint our system provides an interesting example, since the origin of the phase transition is due to a non-standard behaviour of the grand-canonical measures, in particular the lack of a law of large numbers. This leads to a richer behaviour than in previous models, which can be fully understood only by studying the canonical measures as well, which is not the case for zero-range processes with fixed jump rates \cite{stefan2}. The mathematical structure is also different from standard results on systems with bounded Hamiltonians \cite{ellisetal00,touchetteetal04}.  
We also show how our findings can be directly generalized to a process where the jump rates depend on the total number of particles, rather than the size of the lattice. To establish the link between the stationary distribution and dynamics we include a discussion of metastability and the life times of metastable phases, which are compared to Monte Carlo simulation data. Our results can be generalized heuristically to a large class of systems, including models of granular clustering, as is explained in a forthcoming publication \cite{tobepublished}.

The paper is organized as follows. In the next section we introduce the model and show its phase diagram, which summarizes our results. In Section 3 we study canonical and grand-canonical stationary measures and the equivalence of ensembles is discussed in Section 4. In Section 5 we present results on metastability and in Section 6 on the extension to a dependence on the number of particles in the system. In the discussion in Section 7 we give a detailed comparison with previous results.

%\pagebreak
\section{Model and results}

We consider a zero-range process on a translation invariant lattice $\Lambda_L$ of size $|\Lambda_L |=L$. The state space is given by the set of all particle configurations,
\bea\label{statespace}
X_L =\big\{\feta =(\eta_x )_{x\in\Lambda_L} \, :\,\eta_x \in\N\big\}\ ,
\eea
where the number of particles per site can be any non-negative integer number. With rate $g_R (\eta_x )$ one particle leaves site $x\in\Lambda_L$, and jumps to another site $y$ with probability $p(y-x)$. To avoid degeneracies, we require the jump probabilities $\big\{ p(x)\,\big|\, x\in\Lambda_L \big\}$ to be irreducible and of finite range, i.e. $p(x)=0$ if $|x| >C$ for some $C>1$. Under these conditions our main results are independent of the actual choice of $p$. Since they cover the basic novelties of the paper, we restrict ourselves to the jump rates of the form
\bea\label{rates}
g_R (k)=\left\{\bacl c_0 \ &,\ k\leq R \\ c_1 \ &,\ k> R\ea\right.\quad\mbox{for }k\geq 1\ ,\quad g(0)=0\ ,
\eea
where $c_0 >c_1 >0$. The rates are piecewise constant and the location of the jump is given by the parameter $R\geq 0$, which depends on the system size $L$, such that
\bea\label{apar}
R\to\infty\quad\mbox{and}\quad R/L\to a\quad\mbox{as }L\to\infty\ ,
\eea
where $a\geq 0$ is a system parameter. The most interesting case we will consider is $a>0$, but we will also discuss $a=0$ which depends on the asymptotic behaviour of $R$ as $L$ tends to $\infty$. The same model has already been mentioned in \cite{evans00} for fixed $R$. There is no phase transition in this case, but for large $R$ one observes a large crossover, i.e. convergence in the thermodynamic limit is very slow.

The generator of the process is given by
\bea\label{generator}
\mathcal{L}f(\feta )=\sum_{x,y\in\Lambda_L} g_R (\eta_x )\, p(y-x)\,\big( f(\feta^{x,y} -f(\feta )\big)\ .
\eea
It is defined for all continuous cylinder functions $f\in C(X_L )$. Since we define the process only on finite lattices, there are no further restrictions on initial conditions or the domain of the generator as opposed to zero-range processes on infinite lattices (cf. \cite{andjel82}). We do not specify the geometry or the dimension of the lattice, since our main results on the stationary distribution do not depend on these details. The only requirement is that the lattice is translation invariant, or more generally, $\phi_x =const.$ is the only positive solution to the difference equation
\bea\label{difference}
\phi_x =\sum_{y\in\Lambda_L} \phi_y p(x-y)\ .
\eea
Note that no particles are created or annihilated and the number of particles is a conserved quantity. Under our assumptions on $p$ and $g$ there are no other conservation laws that would lead to degeneracies in the time evolution.

\begin{figure}
\begin{center}
\includegraphics[width=0.48\textwidth]{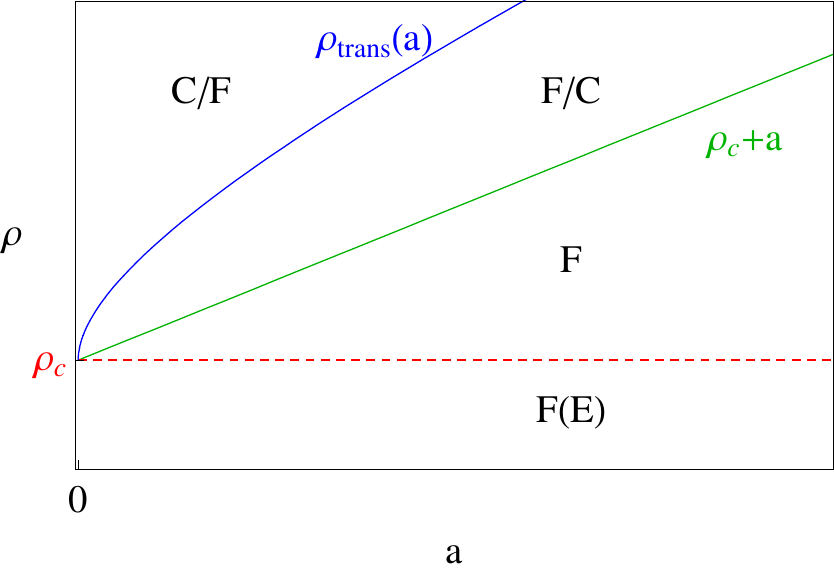}\hfill
\includegraphics[width=0.48\textwidth]{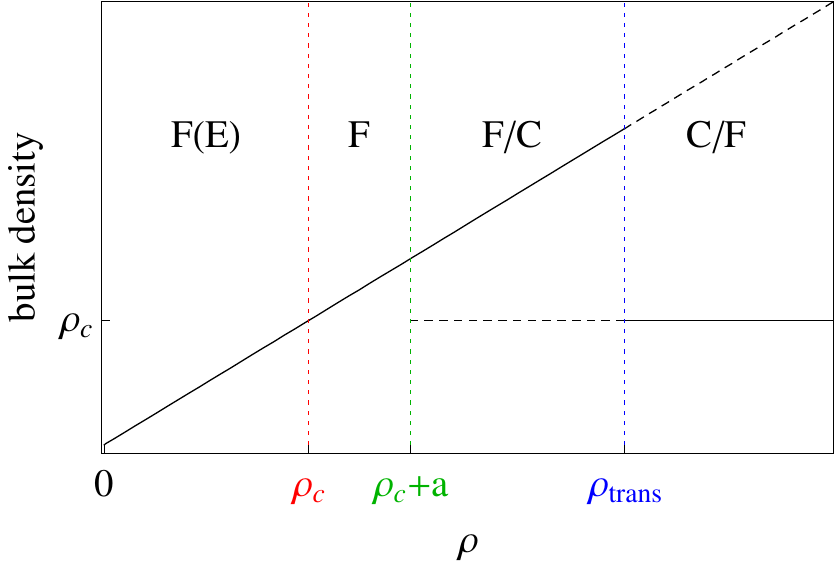}\\
\end{center}
\caption{\label{fig1} Stationary phase diagram for generic values of $c_0 >c_1$.
The four phases $F(E)$ ($\rho\leq\rho_c$), $F$ ($\rho_c <\rho\leq\rho_c +a$), $F/C$ ($\rho_c +a<\rho <\rho_{trans}$) and $C/F$ ($\rho \geq\rho_{trans}$) are explained in the text.\newline
Left: Phase diagram in terms of $a$ (\ref{apar}) and the particle density $\rho$. Right: Background density as a function of $\rho$ for $a=0.5$. Full lines are stable, broken lines metastable.}
\end{figure}

For fixed $L$, also $R$ is a fixed parameter and known results on stationary measures for zero-range processes apply (see e.g.~\cite{evansetal05} and references therein). The stationary weight $w_R^L(\feta)$ for this process is of product form, 
\bel{product}
w_R^L(\feta) = \prod_{x\in\Lambda_L} w_R(\feta_x),
\ee
where the single-site marginal is given by
\bea\label{weight}
w_R (k)=\prod_{i=0}^k g_R^{-1} (i)=\left\{\bacl c_0 ^{-k} \ &,\ k\leq R\\ c_0^{-R} c_1^{R-k}\ &,\ k>R\ea\right.\ .
\eea
Here the empty product (for $k=0$) is understood to be unity.

The results we derive in the following sections are summarized in the stationary phase diagram in Figure \ref{fig1} in terms of the conserved particle density $\rho$ and the parameter $a$ (\ref{apar}). In the fluid phases $F(E)$, $F$ and $F/C$ the stationary measure concentrates on homogeneous configurations with bulk density $\rho$. In phase $F(E)$ for $\rho\leq\rho_c$ the canonical and grand-canonical ensembles are equivalent (see Section 4), and in phase $F/C$ for $\rho_c +a<\rho\leq\rho_{trans}$ there exists an additional metastable condensed phase, which has a lifetime exponential in the system size (see Section 5). Typical condensed configurations have a $\rho$-independent homogeneous bulk distribution with density $\rho_c <\rho$, where the excess particles condense on a single lattice site. In phase $C/F$, i.e. for $\rho >\rho_{trans}$, the condensed phase becomes stable and the corresponding fluid phase metastable. On top of metastability the order parameters are discontinuous as a function of the density $\rho$, and therefore the condensation transition is discontinuous.

\section{Stationary measures}
\subsection{Grand-canonical measures}

Since the state space $X_L$ is discrete we will identify measures $\mu \big(\{ \feta\}\big)$ with their mass functions $\mu (\feta )$ in the following to simplify notation. For each $R$ and $L$ there exists a family of stationary product measures $\nu_{\phi ,R}^L$ with single site marginal
\bea\label{mass}
%\nu^1_{\phi ,R} \big(\{ k\}\big) =
\nu^1_{\phi ,R} (k)=\frac1{z_R (\phi )}\, w_R (k)\,\phi^k\ .
\eea
The marginal is well defined for fugacities $\phi\in [0,c_1 )$, since the tail behaviour of the stationary weight (\ref{weight}) is $w_R (k)\sim c_1^{-k}$ for all fixed $R$. The single site normalization is given by the partition function
\bea\label{zR}
z_R (\phi )=\sum_{k=0}^\infty w_R (k)\,\phi^k =\frac{c_0}{c_0 -\phi}\bigg(1+ \Big(\frac{\phi}{c_0}\Big)^{R+1} \frac{c_0 -c_1 }{c_1 -\phi }\bigg)
\eea
and the expected particle density under the measure $\nu_{\phi ,R}^L$ is given by
\bea\label{rhoR}
\rho_R (\phi )&=&\big\langle\eta_x \big\rangle_{\nu_{\phi ,R}^1} =\phi\,\partial_\phi \big(\log z_R (\phi )\big) =\nonumber\\
&=&\frac{\phi}{c_0 -\phi} +\Big(\frac{\phi}{c_0}\Big)^{R+1} \frac{R+1+\phi /(c_1 -\phi)}{\frac{c_1 -\phi}{c_0 -c_1}+(\phi /c_0)^{R+1}}\ .
\eea
Note that $\rho_R (\phi )$ is strictly increasing in $\phi$ and that for every fixed $R$, $\rho_R (\phi )\to\infty$ as $\phi\to c_1$. So for all densities $\rho\geq 0$ there exists $\phi_R (\rho )$ such that the measure $\nu_{\phi_R (\rho ) ,R}^L$ has density $\rho$, i.e. product measures exist for all densities. But the single site marginals of these measures still depend on $R$ and therefore on the system size $L$. Since $R\to\infty$ as $L\to\infty$, the marginal (\ref{mass}) converges pointwise to a simple geometric distribution, i.e. for each $k\in\N$,
\bea\label{massinfty}
\nu^1_{\phi ,R} (k)\to\nu^1_{\phi ,\infty} (k)=\frac1{z_\infty (\phi )}(\phi /c_0 )^k \quad\mbox{with}\quad z_\infty (\phi )=\frac{c_0}{c_0-\phi}\ .
\eea
This convergence holds for each fixed $\phi <c_1$, but it is not uniform in $\phi$. The limiting product measure $\nu_{\phi ,\infty}$ is defined for all $\phi <c_0$. We denote the particle density with respect to this measure by
\bea\label{rhoinf}
\rho_\infty (\phi ):=\big\langle\eta_x \big\rangle_{\nu_{\phi ,\infty}^1}=\phi\,\partial_\phi \big(\log z_\infty (\phi )\big)=\frac{\phi}{c_0 -\phi} \ ,
\eea
and its inverse is given by
\bea\label{phiinfty}
\phi_\infty (\rho )=c_0 \,\frac{\rho}{1+\rho} \ .
\eea
Since convergence (\ref{massinfty}) only holds for $\phi <c_1$ we define the critical density
\bea\label{rhoc}
\rho_c :=\rho_\infty (c_1 )=\frac{c_1}{c_0 -c_1} <\infty\ .
\eea
Note that with this definition $\phi_\infty (\rho_c )=c_1$. In the following we summarize some straightforward consequences of these definitions.

\begin{proposition}\label{prop1a}
For all $\phi <c_1$, $\nu^L_{\phi ,R} \to\nu_{\phi ,\infty}$ weakly or, equivalently,
\bea\label{weakconv1}
\langle f\rangle_{\nu^L_{\phi ,R}} \to\langle f\rangle_{\nu_{\phi ,\infty}}\quad\mbox{as }L\to\infty\mbox{ for all }f\in C_{0,b} (X)\footnotemark\ ,
\eea\footnotetext{$C_{0,b} (X)$ denotes the set of all bounded, continuous cylinder functions $f:X\to\R$. A cylinder function depends only on the particle configuration on a fixed finite number of lattice sites.}
and $\rho_R (\phi )\to\rho_\infty (\phi )$. For all $\rho\geq 0$ we have
\bea\label{phir}
\phi_R (\rho )\to\left\{\bacl \phi_\infty (\rho)&,\ \rho {<}\rho_c \\ c_1 &,\ \rho {\geq}\rho_c\ea\right. \quad\mbox{and}\quad \nu_{\phi_R (\rho ),R}^L \to\left\{\bacl \nu_{\phi_\infty (\rho ),\infty}&,\ \rho {<}\rho_c \\ \nu_{c_1 ,\infty}&,\ \rho {\geq}\rho_c\ea\right.\ ,
\eea
where $\phi_R$ is the inverse of (\ref{rhoR}) and the second convergence holds in the weak sense as in (\ref{weakconv1}).
\end{proposition}

\noindent\textbf{Proof.} (\ref{massinfty}) implies pointwise convergence of arbitrary $n$-point marginals $\nu_{\phi ,R}^n$ and in general this is equivalent to convergence of expected values of cylinder test functions, as long as they are bounded. This does not directly imply convergence of the unbounded test function $\eta_x$ which yields the density, but $\rho_R (\phi )\to\rho_\infty (\phi )$ follows by direct computation from (\ref{rhoR}).\\
Since $\rho_R (\phi )$ and its inverse are continuous for $\phi <c_1$ or equivalently $\rho <\rho_c$, we have $\phi_R (\rho )\to\phi_\infty (\rho )$. Since $\phi_\infty (\rho )<c_1$ and $z_\infty$ is a continuous function, inserting $\phi_R (\rho )$ in (\ref{zR}) yields as $L\to\infty$
\bea\label{zconv}
z_R \big(\phi_R (\rho )\big) =z_\infty \big(\phi_R (\rho )\big)\bigg(1{+} \Big(\frac{\phi_R (\rho )}{c_0}\Big)^{R+1} \frac{c_0 -c_1 }{c_1 {-}\phi_R (\rho )}\bigg)\to z_\infty \big(\phi_\infty (\rho )\big)\, .
\eea
Therefore, we have pointwise convergence of the marginals as in (\ref{massinfty}) and $\nu_{\phi_R (\rho ),R}^L \to\nu_{\phi_\infty (\rho ),\infty}$ weakly for $\rho <\rho_c$ analogous to above. For $\rho >\rho_c$ to leading order
\bea\label{phirlead}
\phi_R (\rho )\simeq c_1 -\Big(\frac{c_1}{c_0}\Big)^{R/2} \frac{c_1}{\sqrt{z_\infty (c_1)(\rho -\rho_c)}}\to c_1 \quad\mbox{as }L\to\infty\ .
\eea
For $\rho =\rho_c$ the correction has a different power $\big(\frac{c_1}{c_0}\big)^{R/4}$ which leads to the same behaviour as for $\rho >\rho_c$. Inserting in (\ref{zR}) this yields analogous to (\ref{zconv})
\bea\label{zrconv}
z_R \big(\phi_R (\rho )\big)=z_\infty \big(\phi_R (\rho )\big)\bigg( 1+\Big(\frac{c_1}{c_0}\Big)^{R/2}\sqrt{\frac{\rho -\rho_c}{z_\infty (c_1)}}\bigg)\to z_\infty (c_1 )\quad\mbox{as }L\to\infty
\eea
so that $\nu^L_{\phi_R (\rho ),R}\to\nu_{c_1 ,\infty}$ weakly.\hfill $\Box$\\

\noindent Note that by Proposition \ref{prop1a} the density does not converge if $\rho >\rho_c$ since
\bea\label{rhodiv}
\rho_R \big(\phi_R (\rho )\big) =\rho\not\to\rho_c =\rho_\infty (c_1 )\ ,
\eea
and the variance of $\eta_x$ even diverges as
\bea\label{vardiv}
Var (\eta_x )=\phi\,\partial_\phi \rho_R (\phi )\big|_{\phi =\phi_R (\rho )}\simeq 2c_1 \Big(\frac{c_0}{c_1}\Big)^{R/2-1}%(\rho -\rho_c )\, R\ .
\eea
Therefore there is no standard law of large numbers for the measures $\nu^L_{\phi_R (\rho ),R}$ when $\rho >\rho_c$. In particular one can show the following.

\begin{proposition}\label{prop1}
For each $L$ let $\eta_1^L ,\ldots ,\eta_L^L$ be iid random variables with distribution $\nu^1_{\phi_R (\rho ),R}$ and assume that $R\gg\log L$. Then
\bea\label{nlln}
\frac1L \sum_{x\in\Lambda_L} \eta_x^L \to\left\{\bacl \rho\ &,\ \rho <\rho_c \\ \rho_c \ &,\ \rho\geq\rho_c\ea\right.\quad \mbox{almost surely}\ .%\mbox{in distribution}\ ,
\eea
%where $X\sim Poi\big(\frac{\rho -\rho_c}{a}\big)$ is a Poisson random variable.
\end{proposition}

\noindent\textbf{Proof.} see appendix\\

\noindent Note that for $\rho >\rho_c$ (\ref{rhodiv}) holds due to very large values $\eta_x \sim \big(\frac{c_0}{c_1}\big)^{R/2}$ having very small probabilities $\big(\frac{c_1}{c_0}\big)^{R/2}$, which also leads to divergence of the variance (\ref{vardiv}). In turn, the small probabilities lead to almost sure convergence of the sample mean to $\rho_c <\rho$, which is a non-standard strong law of large numbers. The breakdown of the standard strong law coincides with the region of nonequivalence of ensembles, as has been observed also in the context of spin systems \cite{ellisetal00,touchetteetal04}.
%\noindent Note that for $\rho\leq\rho_c$ convergence also holds almost surely for any construction of the $\eta_x^L$ on a common probability space. The interpretation of (\ref{nlln}) is that for $\rho >\rho_c$ a Poisson distributed number of sites contributes of order $a\, L+o(L)$ to the sum. Note that $\langle X\rangle =Var (X)=(\rho -\rho_c )/a$ in accordance with (\ref{rhodiv}) and (\ref{vardiv}), and for $\rho >\rho_c +a$ we have on average at least one site with $\eta_x \simeq aL$.
%This already points to the fact that the behaviour of the system is non-standard for $\rho >\rho_c$.

\begin{figure}
\begin{center}
\includegraphics[width=0.48\textwidth]{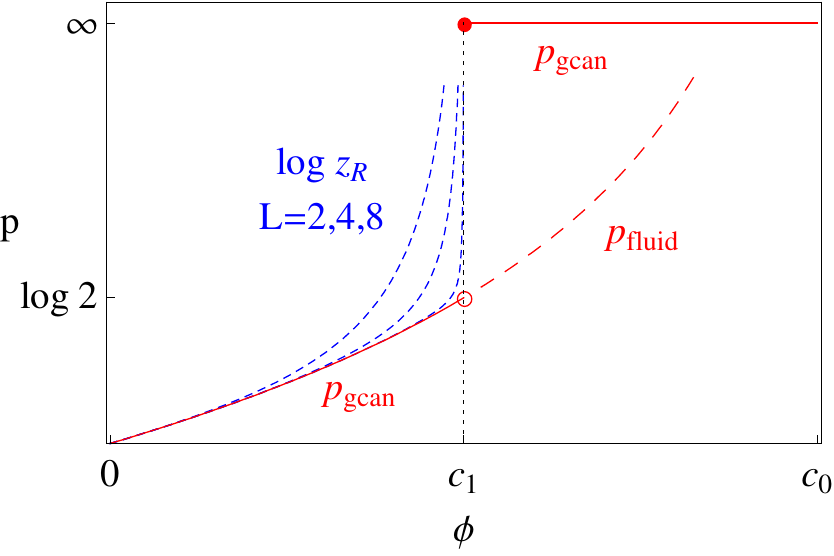}\hfill\includegraphics[width=0.48\textwidth]{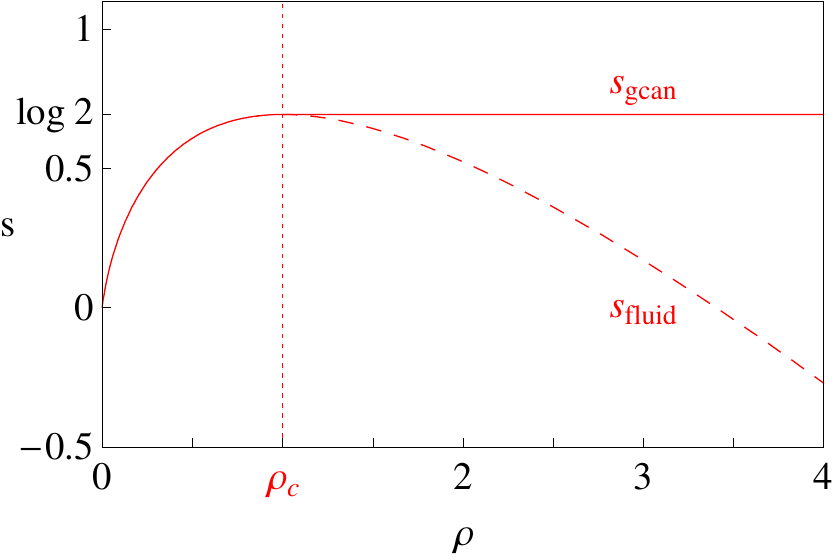}\\
\end{center}
\caption{\label{fig2} Properties of fluid and grand-canonical measures for $c_0 =2$, $c_1 =1$ as given in (\ref{pfluid}) to (\ref{sgcan}). Left: Pressure $p_{gcan}$ (full red line), $p_{fluid}$ (broken red line) and $\log z_R$ for $L=2,4,8$ (dashed blue lines), demonstrating the fast convergence to $p_{gcan}$. Right: Entropy densities $s_{gcan}$ (full red line) and $s_{fluid}$ (broken red line).}
\end{figure}

As noted before, the limiting product measures $\nu_{\phi ,\infty}$ (\ref{massinfty}) exist for all $\phi <c_0$ and for reasons explained below, we call the family of measures $\big\{ \nu_{\phi_\infty (\rho ),\infty}\, :\,\rho\geq 0\big\}$ the \textit{fluid phase}. The pressure of the fluid phase is given by
\bea\label{pfluid}
p_{fluid} (\phi ):=\lim_{L\to\infty} \frac1L\log z_\infty^L (\phi )=\log z_\infty (\phi )=\log\frac{c_0}{c_0 -\phi}
\eea
and we define the entropy density by the negative Legendre transform
\bea\label{sfluid}
s_{fluid} (\rho )&:=&-\sup_{\phi\geq 0} \big(\rho\,\log\phi -p_{fluid} (\phi )\big) =p\big(\phi_\infty (\rho )\big) -\rho\log\phi_\infty (\rho )=\nonumber\\
&=&(1+\rho )\log (1+\rho )-\rho (\log c_0 +\log\rho )\ ,
\eea
where the supremum is attained for $\phi =\phi_\infty (\rho )$ (\ref{phiinfty}).
Note that the fluid pressure and entropy density are different from the grand-canonical quantities, because $z_R (\phi )=\infty$ for $\phi\geq c_1$ (\ref{zR}). This yields
\bea\label{pgcan}
p_{gcan} (\phi ):=\lim_{L\to\infty} \frac1L\log z_R^L (\phi )=\left\{\bacl p_{fluid} (\phi )\ &,\ \phi <c_1 \\ \infty\ &,\ \phi \geq c_1 \ea\right.
\eea
and the negative Legendre transform of the pressure is given by
\bea\label{sgcan}
s_{gcan} (\rho )=\left\{\bacl s_{fluid} (\rho )\ &,\ \rho\leq\rho_c \\ s_{fluid} (\rho_c )-(\rho -\rho_c )\log c_1 \ &,\ \rho >\rho_c \ea\right.\ .
\eea
Note that the Legendre transform of the pressure is usually called the free energy density. In thermodynamics, the free energy $F$ is related to the entropy $S$ via $F=U-TS$, where $U$ is the internal energy and $T$ the temperature. Since there is no energy and temperature in our case, we define the entropy density as the negative free energy density. The functions (\ref{pfluid}) to (\ref{sgcan}) are illustrated in Figure \ref{fig2}. In analogy to previous results \cite{stefan,stefan2} we expect a condensation transition for $\rho >\rho_c$. But the non-standard behaviour of the grand-canonical measures, in particular the lack of a law of large numbers (\ref{nlln}), will lead to a richer behaviour than in previous studies, which can be fully understood only in the context of the equivalence of ensembles. In particular, the grand-canonical approach alone does not provide a complete picture of the phase transition.

\subsection{Canonical measures}

The canonical measures are given by
\bea
\pi_{L,N} :=\nu_{\phi ,R}^L (\, .\, |\, \Sigma_L =N)\quad\mbox{where}\quad \Sigma_L (\feta ):=\sum_{x\in\Lambda_L} \eta_x\ ,
\eea
i.e. they are given by a grand-canonical measure conditioned on a fixed number $N$ of particles. Their mass functions are independent of $\phi$ and given 
in terms of the stationary weights (\ref{product}) by
\bea\label{canens}
\pi_{L,N} (\feta )=\frac{1}{Z_{L,N}}\, w_R^L (\feta )\,\delta (\Sigma_L (\feta ),N)\ ,
\eea
concentrating on configurations
\bea
X_{L,N} =\big\{\feta\in X_L \,\big|\,\Sigma_L (\feta )=N\big\}\ .
\eea
The partition function is now given by the finite sum
\bea
Z_{L,N} =w_R^L (X_{L,N} )=\sum_{\feta\in X_{L,N}} w_R^L (\feta )\ .
\eea
In the following we analyze the limiting behaviour of this quantity.
In the discussion configurations with many particles on a small number
of sites turn out to play an important role. Therefore we define the 
disjoint sets of configurations
\bea\label{decompose}
X_{L,N}^m =\big\{\feta\in X_{L,N}\,\big|\, \eta_x >R\mbox{ for exactly }m\mbox{ sites }x\in\Lambda_L \big\}\ 
\eea
with more than $R$ particles on
exactly $m$ sites.

\begin{theorem}\label{theo1}
Suppose $R\gg\log L$, i.e. $\frac{\log L}{R}\to 0$ as $L\to\infty$. Then the limit
\bea\label{scan1}
s_{can} (\rho ):=\lim_{L\to\infty} \frac1L\log Z_{L,N} \ ,\quad\mbox{where }N/L\to\rho\ ,
\eea
exists and is called the \textit{canonical entropy density}. It is given by
\bea\label{scan2}
s_{can} (\rho )=\left\{\bacl s_{fluid} (\rho )\ &,\ \rho \leq\rho_{trans}\\ s_{fluid} (\rho_c )+s_{cond} (\rho ,\rho_c)\ &,\ \rho >\rho_{trans}\ea\right.\ ,
\eea
where
\bea\label{scond}
s_{cond} (\rho ,\rho_c )=\lim_{L\to\infty} \frac1L \log w_R \big( (\rho -\rho_c )L\big)\ .
\eea
The transition density $\rho_{trans} (a)$ is given by the unique solution of
\bea\label{acon}
a=\Big( s_{fluid} (\rho_c )-(\rho -\rho_c )\log c_1 -s_{fluid} (\rho )\Big)\Big/\log\frac{c_0}{c_1} \ ,
\eea
where $\rho_{trans} (a)\geq\rho_c +a$ with equality if and only if $a=0$.
\end{theorem}

\noindent Note that with (\ref{scond}) and (\ref{weight}) the contribution of the condensate to the canonical entropy is given by
\bea
s_{cond} (\rho ,\rho_c )=-(\rho -\rho_c )\log c_1 -a\log\frac{c_0}{c_1}\ .
\eea
As a special case, taking $a=0$ we have $\rho_{trans} =\rho_c$ as the unique solution of (\ref{acon}), and comparing (\ref{scan2}) with (\ref{sgcan}) yields
\bea\label{a0equi}
s_{can} (\rho )=s_{gcan} (\rho )\quad\mbox{for all }\rho\geq 0\ .
\eea
On the other hand, both entropies are different whenever $a>0$.\\

\noindent\textbf{Proof of Theorem \ref{theo1}.} Using (\ref{decompose}) we 
decompose the state space $X_{L,N} =\bigcup_{m=0}^M X_{L,N}^m$. The maximal number $M$ of sites containing more than $R$ particles is certainly bounded by $M=:\lceil N/R\rceil$. Notice that $M\to\rho /a\in (0,\infty ]$ as $L\to\infty$, and in particular $M/L\to 0$. We can estimate the 
number of ``uncondensed'' configurations where no site has more than $R$ particles
by the following Lemma, which is proved in the appendix.
\begin{lemma}\label{lemma1}
For all $L,N\geq 1$ and $M$ as above we have
\bea
\frac{|X_{L,N}|}{1+{L+M\choose M}\big/ (L-M)^R}\leq |X_{L,N}^0 |\leq |X_{L,N} |\ .
\eea
This includes for all $\rho\geq 0$ and $N/L\to\rho$
\bea
\lim_{L\to\infty} \frac1L\log |X_{L,N}^0 |=\lim_{L\to\infty} \frac1L\log |X_{L,N} |=\chi (\rho )\ ,
\eea
where $\chi (\rho ):=(1+\rho )\log (1+\rho )-\rho\log\rho$.\\
Furthermore, if $R\gg\sqrt{L}$, then $\lim\limits_{L\to\infty} |X_{L,N}^0 |/|X_{L,N} |=1$ for all $\rho\geq 0$.
\end{lemma}

\noindent Now we split the partition function accordingly
\bea
Z_{L,N} =\sum_{m=0}^M Z_{L,N}^m \ ,\quad\mbox{where}\quad Z_{L,N}^m =w_R^L (X_{L,N}^m )\ .
\eea
For the term $m=0$ we get with Lemma \ref{lemma1}
\bea\label{n0term}
\frac1L\log Z_{L,N}^0 =\frac1L\log \big( c_0^{-N} |X_{L,N}^0 |\big)\to \chi (\rho )-\rho\log c_0 =s_{fluid} (\rho )\ .
\eea
The contributions of the other terms are given by
\bea\label{others}
Z_{L,N}^m {=}{L\choose m} c_0^{-mR}\!\!\!\sum_{k=m(R+1)}^N \!\!\! c_0^{-(N-k)}\, c_1^{-(k-mR)} \,\big| X_{L-m,N-k}^0 \big| {k{-}mR{-}1\choose m{-}1} .
\eea
Here we have chosen $m$ sites on which we distribute $k$ particles such that each site contains at least $R+1$ particles, giving rise to the first and last combinatorial factor. The $N-k$ remaining particles are distributed on $L-m$ sites such that none contains more than $R$ particles. The sum can be approximated by an integral and evaluated by the saddle point method. The saddle point equation reads
\bea\label{saddle}
\log\frac{c_0}{c_1} -\frac{L}{L-m}\ \chi'\Big(\frac{N-k}{L-m}\Big) +\log\frac{k-mR-1}{k-m(R+1)}=0\ .
\eea
This has a solution if and only if
\bea
N-(L-m)\rho_c \geq m(R+1)\quad\mbox{or, equivalently}\quad\rho \geq\rho_c +m\, a\ .
\eea
In this case, to leading order the solution to (\ref{saddle}) is given by
\bea
k\simeq N-(L-m)\rho_c
\eea
where we have used that $m/L\to 0$ for all $m\leq M$. On the other hand, for $\rho <\rho_c +m\, a$ the sum in (\ref{others}) is maximized for the boundary value $k=m(R+1)$. We get in leading exponential order
\bea\label{zmbehav}
Z_{L,N}^m \simeq\left\{\bacl 0\ &,\ \rho\leq m\, a\\ {L\choose m}\big(\frac{c_0}{c_1}\big)^m c_0^{-N} e^{(L-m)\chi (\rho -ma)}\ &,\ m\, a{<}\rho {<}\rho_c {+}m\, a\\ {L\choose m}\big(\frac{c_1}{c_0}\big)^{mR+(L-m)\rho_c} c_1^{-N} e^{(L-m)\chi (\rho_c )}\ &,\ \rho\geq\rho_c +m\, a\ea\right.\ .
\eea
For $\rho >\rho_c +a$ we get a rough estimate by adding both cases,
\bea\label{n2sum}
\sum_{m=2}^M Z_{L,N}^m \leq Z_{L,N}^1{L\choose M} \bigg(\frac{M}{L}\Big(\frac{c_1}{c_0}\Big)^{N-M-R-(L-1)\rho_c}  {+}\Big(\frac{c_1}{c_0}\Big)^R C L^{M-2} \bigg)\, ,
\eea
where $C=\exp\Big( (c_0 /c_1)^{\rho_c} e^{-\chi (\rho_c )}\Big)$. Now, to leading order
\bea\label{n2all}
\lefteqn{\frac1L\log \Bigg( {L\choose M} \Big(\frac{c_1}{c_0}\Big)^R C\, L^{M-2}\Bigg) \simeq -\frac{M}{L}\Big( 1+\log\frac{M}{L}+\frac{M}{L}\Big)-\frac{\log L}{L}}\nonumber\\
& &\quad -\frac{R}{L}\log\frac{c_0}{c_1}+\frac{M}{L}\log L\ \to\  -a\log\frac{c_0}{c_1} \leq 0\quad\mbox{as }L\to\infty\ ,
\eea
since $M/L\to 0$, $R/L\to a\geq 0$. This holds only if $M\ll L/\log L$ or, equivalently, $R\gg\log L$. Since $\rho >\rho_c +a$ the first summand on the right-hand side of (\ref{n2sum}) vanishes with an analogous argument. Therefore
\bea
\frac1L\log\bigg(1+\sum_{m=2}^M Z_{L,N}^m \Big/ Z_{L,N}^1\bigg) \to 0\ ,
\eea
and the only exponential contribution to (\ref{n2sum}) is given by $Z_{L,N}^1$. Thus we have, using (\ref{zmbehav}),
\bea\label{nterm}
\lefteqn{\lim_{L\to\infty}\frac1L\log\sum_{m=1}^M Z_{L,N}^m =\lim_{L\to\infty}\frac1L\log Z_{L,N}^1 =}\nonumber\\
& &\quad =\lim_{L\to\infty}\frac1L\log\bigg( L\Big(\frac{c_1}{c_0}\Big)^{R+(L-1)\rho_c} c_1^{-N} \big| X^0_{L-1,(L-1)\rho_c}\big|\bigg) =\nonumber\\
& &\quad =(a+\rho_c )\log\frac{c_1}{c_0} -\rho\log c_1 +\chi (\rho_c )=s_{fluid} (\rho_c )+s_{cond} (\rho ,\rho_c)\ .
\eea
This is a linear function in $\rho$ with the same slope $-\log c_1$ as $s_{gcan} (\rho )$ (\ref{sgcan}). Note that for $\rho\to\infty$ the first term (\ref{n0term}) behaves as
\bea
s_{fluid} (\rho )\simeq -\rho\log c_0 +\log (1+\rho)+1\ .
\eea
Therefore, whereas for small $\rho$ (\ref{n0term}) dominates the partition function, (\ref{nterm}) dominates for large $\rho$, since it has larger asymptotic slope $-\log c_1 >-\log c_0$. The transition density $\rho_{trans}$ as a function of $a$ is found by equating both contributions which leads directly to (\ref{acon}). Differentiating the right-hand side of this equation yields
\bea
a'(\rho )=1-\log\frac{1+\rho}{\rho}\Big/\log\frac{c_0}{c_1}\ .
\eea
Thus $a'(\rho_c) =0$ and $a'(\rho )\in (0,1)$ for all $\rho >\rho_c$. Since also $a(\rho_c )=0$, (\ref{acon}) has a unique solution $\rho_{trans} (a)\geq\rho_c$ for all $a\geq 0$. Further we have
\bea
\rho '_{trans} (a)=\frac1{a'(\rho_{trans} (a) )} >1\quad\mbox{for all}\quad\rho\geq\rho_c \ ,
\eea
and thus $\rho_{trans} (a)\geq\rho_c +a$ with equality if and only if $a=0$.\hfill $\Box$\\

\noindent Now, if $a>0$ then $M$ as defined after (\ref{decompose}) is bounded and converges to $\rho /a$, and thus (\ref{n2sum}) implies that
\bea
\sum_{m=2}^M Z_{L,N}^m \Big/ Z_{L,N}^1 \to 0\quad\mbox{as}\quad L\to\infty\ .
\eea
This is significantly stronger than (\ref{n2all}) and it is easy to see that it still holds for $a=0$, as long as $R\gg\sqrt{L\log L}$. Thus for $L\to\infty$ the canonical measure concentrates on certain parts of the state space, and from the proof of Theorem \ref{theo1} (\ref{n2sum}) the rate of convergence is faster than polynomial in $L$. Therefore we can immediately deduce the following.

\begin{corollary}\label{corr}
For $R\gg\sqrt{L\log L}$ we have
\bea
\rho <\rho_{trans}\quad &\Rightarrow &\quad\pi_{L,N} \big( X_{L,N}^0\big)\to 1\ ,\quad L^n \pi_{L,N} \big( X_{L,N} \setminus X_{L,N}^0\big)\to 0\ ,\nonumber\\
\rho >\rho_{trans}\quad &\Rightarrow &\quad\pi_{L,N} \big( X_{L,N}^1\big)\to 1\ ,\quad L^n \pi_{L,N} \big( X_{L,N} \setminus X_{L,N}^1\big)\to 0\ ,
\eea
for all $n\in\N$ as $L\to\infty$ and $N/L\to\rho$.
\end{corollary}

\noindent This implies in analogy to (\ref{scan2}), that for $\rho >\rho_{trans}$ a typical configuration consists of a homogeneous background with density $\rho_c$ and the $(\rho -\rho_c )L$ excess particles concentrate in a \textit{single} lattice site. We expect this kind of behaviour actually already for $R\gg\log L$, since $w_R$ has an exponential tail and maximal fluctuations under $w_R^L$ in the occupation number are of order $\log L$. Our estimates are not strong enough to deduce this, but we are primarily interested in $a>0$, which is covered by the above result. The same is true for the last statement of Lemma \ref{lemma1}.\\
For $\rho =\rho_{trans}$ the contributions of condensed and fluid configurations to the canonical entropy are equal (\ref{scan2}). This is true on the exponential scale and to deduce the behaviour on the transition line we need a finer estimate, given in the following Theorem.

\begin{theorem}\label{theo1b}
For $N/L\to\rho_{trans}$ and $a>0$ we have
\bea
w_R^L (X_{L,N}^1)/w_R^L (X_{L,N}^0)= O(L^{3/2})\to\infty\quad\mbox{as }L\to\infty\ ,
\eea
which implies $\pi_{L,N} (X_{L,N}^1 )\to 1$.
\end{theorem}

\noindent So in case of a discontinuous transition (i.e. $a>0$) the transition line belongs to the condensed phase $C/F$. For $a=0$ the transition is continuous and therefore $\rho =\rho_c$ belongs to the fluid phase $F(E)$.\\

\noindent\textbf{Proof.} According to (\ref{others}) in the proof of Theorem \ref{theo1},
\bea
w_R^L (X_{L,N}^1) &=&L\, c_0^{-R}\sum_{k=R+1}^N c_0^{-(N-k)}\, c_1^{-(k-R)} \,\big| X_{L-1,N-k}^0 \big| =\nonumber\\
&=& L\Big(\frac{c_1}{c_0}\Big)^{R+\rho_c (L-1)} c_1^{-N} \big| X_{L-1,\rho_c (L-1)}^0 \big|\,\big(1+o(1)\big)\nonumber\\
& &\int_{R+1}^N \exp\bigg(\frac12 \,\chi''(\rho_c )\frac{L}{(L-1)^2}(k-\bar k)^2 \bigg) dk
\eea
where $\bar k=N-(L-1)\rho_c +o(L)$ is the solution to the saddle point equation (\ref{saddle}). In addition to the proof of Theorem \ref{theo1} we consider the next order of the expansion to get the correct asymptotic behaviour. Since with Lemma~\ref{lemma1}, $\chi''(\rho )=-\frac{1}{\rho (1+\rho )}<0$ for all $\rho >0$ and $\bar k\in (R+1,N)$, the asymptotic behaviour of the Gaussian integral with variance $\sigma^2 =-L/\chi''(\rho_c )(1+o(1))$ is given by its normalization and we get
\bea
\lefteqn{w_R^L (X_{L,N}^1) =}\nonumber\\
& &=L^{3/2} \Big(\frac{c_1}{c_0}\Big)^{R+\rho_c (L-1)} c_1^{-N} \big| X_{L-1,\rho_c (L-1)}^0 \big|\,\sqrt{2\pi \rho_c (1+\rho_c )}\big(1+o(1)\big)\ .
\eea
With $C=\sqrt{2\pi \rho_c (1+\rho_c )}$ this leads to
\bea
\frac{w_R^L (X_{L,N}^1)}{w_R^L (X_{L,N}^0)} &=&CL^{3/2} \frac{\big(\frac{c_1}{c_0}\big)^{R+\rho_c L} c_1^{-N} \big| X_{L-1,\rho_c (L-1)}^0 \big|}{c_0^{-N} \big| X_{L,N}^0 \big|} \big(1+o(1)\big) =\nonumber\\
&=&CL^{3/2} \frac{\big(\frac{c_1}{c_0}\big)^{R+\rho_c L} c_1^{-N} {(L-1)(1+\rho_c )-1\choose L-2}}{c_0^{-N} {L+N-1\choose L-1}} \big(1+o(1)\big)\ ,
\eea
where we have used the third statement of Lemma \ref{lemma1} that holds for $a>0$. We use Stirling's formula for the binomial coefficients and note that due to Theorem \ref{theo1} the exponential terms in the ratio vanish, which leaves us with
\bea
\frac{w_R^L (X_{L,N}^1)}{w_R^L (X_{L,N}^0)} =CL^{3/2} \big( 1+o(1)\big)\to\infty\quad\mbox{as }L\to\infty, N/L\to\rho_{trans}\ .
\eea
Together with Theorem \ref{theo1} this implies that
\bea
w_R^L (X_{L,N}\setminus X_{L,N}^1)\to 0\quad\mbox{as }L\to\infty, N/L\to\rho_{trans}\ ,
\eea
which implies the last statement of the Theorem.\hfill $\Box$\\

\section{Equivalence of ensembles}
\subsection{Specific relative entropy}

\begin{table}
\begin{center}
  \begin{tabular}{c||c|c}
   & canonical entropy & grand-canonical entropy\\
   phase & $s_{can} (\rho )$ & $s_{gcan} (\rho )$\\[2mm]
   \hline\hline &\\
   F(E) & $s_{fluid} (\rho )$ & $s_{fluid} (\rho )$\\[2mm]
   \hline &\\
   F, F/C & $s_{fluid} (\rho )$ & \\[2mm]
   C/F & $s_{fluid} (\rho_c ) {-}(\rho {-}\rho_c)\log c_1 {-}a\log\frac{c_0}{c_1}$ & \raisebox{3mm}[0mm]{$s_{fluid} (\rho_c ) {-}(\rho {-}\rho_c)\log c_1$}
  \end{tabular}
\end{center}
\caption{\label{table1}
Summary of the results of Section 3: Comparison between canonical and grand-canonical entropy density. Equivalence of ensembles holds only in phase $F(E)$.
}
\end{table}

\begin{figure}
\begin{center}
\includegraphics[width=0.48\textwidth]{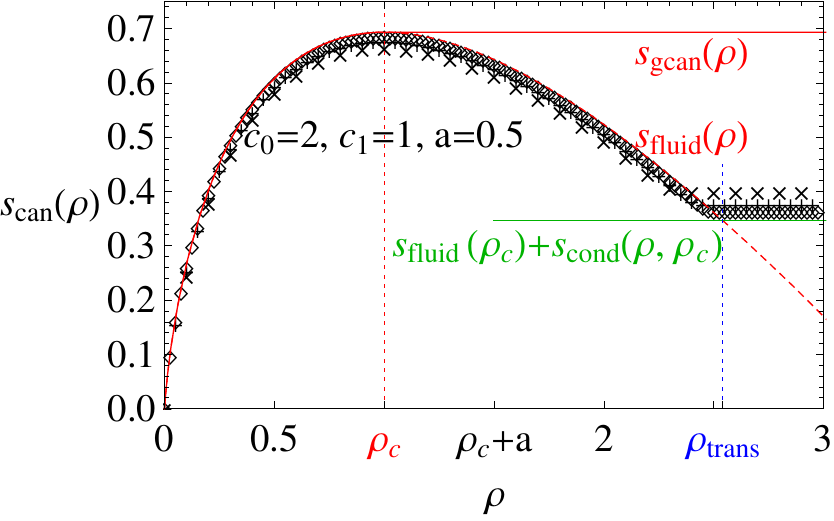}\hfill
\includegraphics[width=0.48\textwidth]{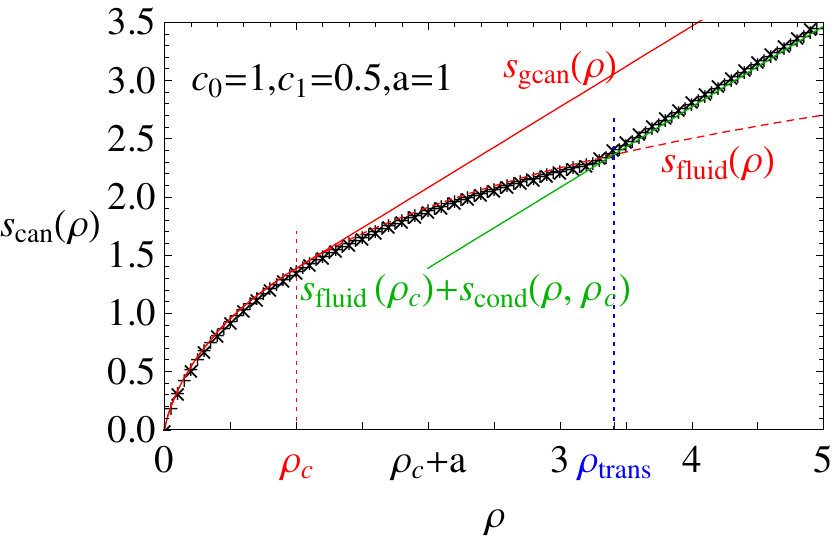}\\[4mm]
\includegraphics[width=0.5\textwidth]{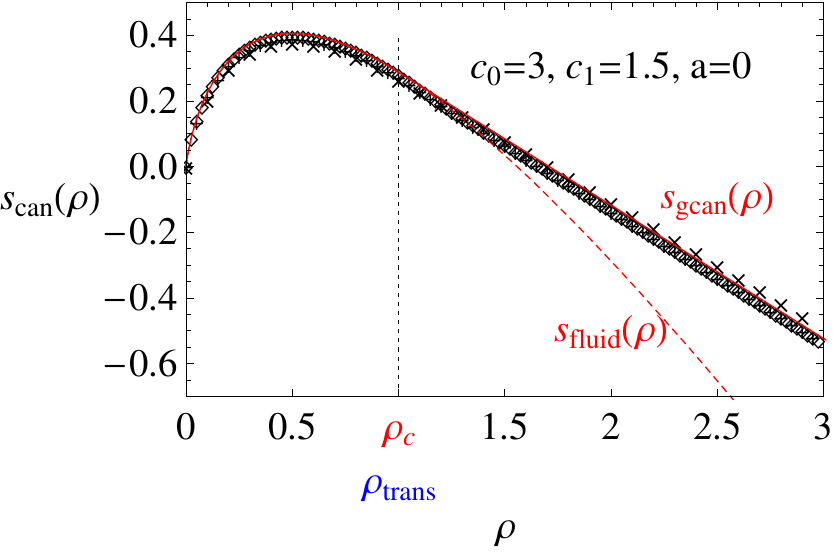}
\end{center}
\caption{\label{fig3}
Canonical entropy density $s_{can} (\rho )$ for various values of $c_0$, $c_1$ and $a$. Data points are calculated numerically according to (\ref{recursion}) with $L=100\ (\times )$, $200\ (+)$, $400\ (\Diamond)$, and show good agreement with the theoretical predictions for the thermodynamic limit (see Table \ref{table1}).}
\end{figure}

In Table \ref{table1} we summarize the results of the previous section in connection with the phase diagram shown in Figure \ref{fig1}. In particular, for $a=0$ the phases $F$ and $F/C$ are empty since $\rho_c =\rho_{trans}$, and we have $s_{can} (\rho )=s_{gcan} (\rho )$ for all $\rho\geq 0$ as noted already in (\ref{a0equi}). This implies that the canonical entropy density is concave and the condensation transition is continuous. On the other hand, for $a>0$ we have equivalence of ensembles only in phase $F(E)$, the canonical entropy density is non-concave, and the transition is discontinuous. These results concern equivalence of ensembles in terms of convergence of entropies of the canonical and the grand-canonical measure. In Figure \ref{fig3} they are illustrated by numerical calculations of the canonical entropy density using the recursion relation
\bea\label{recursion}
Z_{L,N} =\sum_{k=0}^N w_R (k) Z_{L-1,N-k} \ .
\eea

\noindent As can be seen, the grand-canonical entropy density is equal to the concave hull of $s_{can } (\rho )$ which itself is not concave for $a>0$. The canonical entropy density further coincides with the one of the fluid phase, up to the point when it becomes metastable and the condensed phase becomes stable. This point has been derived exactly by studying the dominating terms in the canonical partition function.

We can make a connection to other formulations of the equivalence of ensembles, using the specific relative entropy
\bea\label{relent}
h(\pi_{L,N} ,\nu_{\phi ,R}^L ) :=\frac1L\, H(\pi_{L,N} ,\nu_{\phi ,R}^L ) =\frac1L\,\Big\langle\log\frac{\pi_{L,N}(\feta )}{\nu_{\phi ,R}^L (\feta )} \Big\rangle_{\pi_{L,N}}\ .
\eea
With the identity $\pi_{L,N}^L =\nu_{\phi ,R}^L (\, .\, |\, \Sigma_L =N)$, this can be expressed in two useful forms,
\bea\label{relent2}
h(\pi_{L,N} ,\nu_{\phi ,R}^L )&=&-\frac1L \log\nu_{\phi ,R}^L \big(\Sigma_L =N \big) =\nonumber\\
&=&\log z_R (\phi )-\frac{N}{L}\,\log\phi -\frac1L\,\log Z_{L,N} \ .
\eea
The derivation of these expressions is straightforward, see e.g.~\cite{stefan}. The following is a direct consequence of our results on the canonical measure in Theorem \ref{theo1}.

\begin{corollary}\label{theo2a}
Choosing $\phi =\phi_R (\rho )$ according to (\ref{phir}) we get for all $\rho\geq 0$
\bea\label{relent3}
h(\pi_{L,N} ,\nu_{\phi_R (\rho ) ,R}^L )\to s_{gcan} (\rho )-s_{can} (\rho )\ .
\eea
\end{corollary}

\noindent\textbf{Proof.} We use the second expression in (\ref{relent2}) for the specific relative entropy. Choosing $\phi =\phi_R (\rho )$, the first two terms converge
\bea
\log z_R (\phi_R (\rho ) )-\log\phi_R (\rho )\,\frac{N}{L}\to s_{gcan} (\rho )
\eea
to the grand-canonical entropy density (\ref{sgcan}), since with Proposition \ref{prop1a}, analogous to (\ref{zconv}) and (\ref{zrconv})
\bea
z_R \big(\phi_R (\rho )\big)\to \left\{\bacl z_\infty \big(\phi_\infty (\rho )\big)\ &,\mbox{ for }\rho <\rho_c \\  z_\infty (c_1 )\ &,\mbox{ for }\rho \geq\rho_c \ea\right.\ .
\eea
Convergence of the third term in (\ref{relent2}) has been shown in Theorem \ref{theo1}, which finishes the proof.\hfill $\Box$\\
\\
We can read from Table \ref{table1} that
\bea
s_{gcan} (\rho ){-}s_{can} (\rho )=\left\{\bacl 0&,\, \rho\leq\rho_c \\ \!\!\ s_{fluid} (\rho_c ){-}s_{fluid} (\rho ) {-}(\rho {-}\rho_c )\log c_1 &,\, \rho_c {<}\rho {<}\rho_{trans} \\ a\log (c_0 /c_1 )&,\, \rho \geq\rho_{trans}\ea\right. .
\eea
In particular, for $a=0$ we have $\rho_c =\rho_{trans}$ and
\bea
h(\pi_{L,N} ,\nu_{\phi_R (\rho ) ,R}^L )\to 0\quad\mbox{for all }\rho\geq 0\ ,
\eea
whereas for $a>0$ this holds only for $\rho\leq\rho_c$. By a standard result \cite{csiszar84}, convergence in specific relative entropy implies weak convergence, i.e. convergence of expectations of bounded cylinder test functions $f\in C_{0,b} (X)$,
\bea\label{weakconv}
\Big|\langle f\rangle_{\pi_{L,N}} -\langle f\rangle_{\nu_{\phi_R (\rho ) ,R}^L}\Big|\to 0\quad\mbox{as }L\to\infty\ ,N/L\to\rho\ .
\eea
This is another formulation of the equivalence of ensembles.

Furthermore, we can compare the canonical measures with the expected fluid measures for the background.

\begin{theorem}\label{theo2}
Let $a>0$. Choosing $\phi =\phi_\infty (\rho )$ according to (\ref{phiinfty}) we get
\bea\label{fluidlim}
h(\pi_{L,N} ,\nu_{\phi_\infty (\rho ),\infty}^L )\to s_{fluid} (\rho )-s_{can} (\rho )=0\quad\mbox{for }0\leq\rho <\rho_{trans}\ ,
\eea
whereas for $\phi =c_1$
\bea\label{condlim}
h(\pi_{L,N} ,\nu_{c_1 ,\infty}^L )\to (\rho -\rho_c )\log\frac{c_0}{c_1} >0\quad\mbox{for }\rho \geq\rho_{trans}\ .
\eea
Now let $a=0$ and $R\gg\sqrt{L\log L}$. We have $\rho_{trans} =\rho_c$ and (\ref{fluidlim}) holds for $0\leq\rho\leq\rho_c$,  (\ref{condlim}) for $\rho >\rho_c$. 
\end{theorem}

\noindent\textbf{Proof.} According to the definition (\ref{relent}) we have
\bea\label{zwerg}
\lefteqn{h(\pi_{L,N} ,\nu_{\phi_\infty (\rho ),\infty}^L )=\frac1L \sum_{\feta\in X_{L,N}} \pi_{L,N} (\feta)\log \frac{\pi_{L,N} (\feta)}{\nu_{\phi_\infty (\rho ),\infty}^L (\feta )} =p_{fluid} \big(\phi_\infty (\rho )\big)}\nonumber\\
& &\quad -\frac{N}{L}\,\log\frac{\phi_\infty (\rho )}{c_0}-\frac1L\,\log Z_{L,N} +\frac1L \sum_{\feta\in X_{L,N}} \pi_{L,N} (\feta)\log w_R^L (\feta )\ ,
\eea
where we have used the definitions (\ref{massinfty}) and (\ref{canens}),
\bea
\nu^1_{\phi ,\infty} (k)=\frac1{z_\infty (\phi )}(\phi /c_0 )^k \ ,\quad \pi_{L,N} (\feta )=\frac{1}{Z_{L,N}}\, w_R^L (\feta )\,\delta (\Sigma_L (\feta ),N)\ .
\eea
Splitting the last term of (\ref{zwerg}) and using Corollary \ref{corr} we see that
\bea
\frac1L \sum_{\feta\in X^0_{L,N}} \!\!\!\pi_{L,N} (\feta)\log c_0^{-N} +\frac1L \!\!\!\sum_{\feta\in X_{L,N}\setminus X_{L,N}^0} \!\!\!\!\!\!\!\!\!\!\pi_{L,N} (\feta)\log w_R^L (\feta )\to {-}\rho\log c_0 \, ,
\eea
as $L\to\infty$, $N/L\to\rho$, as long as $\rho <\rho_{trans}$. Therefore, with definition (\ref{sfluid}),
\bea
h(\pi_{L,N} ,\nu_{\phi_\infty (\rho ),\infty}^L )\to s_{fluid} (\rho )-s_{can} (\rho )=0\quad\mbox{for }\rho <\rho_{trans}\ .
\eea
This holds for all $a\geq 0$ as long as $R\gg\sqrt{L\log L}$. 
For $\rho >\rho_{trans}$ we also use (\ref{zwerg}) where $\phi_\infty (\rho )$ is replaced by $c_1$. Again with Corollary \ref{corr} the main contribution to the last term comes now from $\feta\in X^1_{L,N}$. The sum can be computed by the saddle point method analogous to the proof of Theorem \ref{theo1} and we get
\bea
\frac1L \sum_{\feta\in X^1_{L,N}} \pi_{L,N} (\feta)\log w_R^L (\feta)\to a\log\frac{c_1}{c_0}-\rho_c \log c_0 -(\rho -\rho_c )\log c_1 \ .
\eea
The same also holds for $\rho =\rho_{trans}$, since with Theorem \ref{theo1b} $\pi_{L,N}$ concentrates on $X_{L,N}^1$ also in this case. The first terms in (\ref{zwerg}) are now
\bea
p_{fluid} \big( c_1\big) -\frac{N}{L}\,\log\frac{c_1}{c_0} \to s_{fluid} (\rho_c )+\rho\log c_0 -(\rho -\rho_c )\log c_1 \ ,
\eea
and together with the behaviour of $s_{can}$ from Theorem \ref{theo1} we get for $\rho \geq\rho_{trans}$
\bea
h(\pi_{L,N} ,\nu_{c_1 ,\infty}^L )\to (\rho -\rho_c )\log\frac{c_0}{c_1} >0\ ,
\eea
finishing the proof of Theorem \ref{theo2}. Note that $a=0$ is included as a special case in the above derivation as long as $R\gg\sqrt{L\log L}$.\hfill $\Box$\\
\\
(\ref{fluidlim}) allows us to identify the limit measure and we have
\bea
\langle f\rangle_{\pi_{L,N}}\to \langle f\rangle_{\nu_{\phi_\infty (\rho ),\infty}}\quad\mbox{as }L\to\infty\ ,\quad N/L\to\rho\ .
\eea
As a direct consequence of the relative entropy inequality (\cite{csiszar75}, Lemma 3.1), this holds not only for bounded cylinder test functions $f$, but for the larger class with $\langle e^{\epsilon f}\rangle_{\nu_{\phi_\infty (\rho ),\infty}} <\infty$ for some $\epsilon >0$. Since the fluid measures have finite exponential moments, this includes local occupation numbers $f(\feta )=\eta_x$, which are unbounded. This ensures convergence of densities for $\rho <\rho_{trans}$ ($\rho\leq\rho_c$ for $a=0$), i.e. in the fluid phases $F(E)$, $F$ and $F/C$.

  \subsection{The condensed phase}
(\ref{condlim}) may suggest that the limiting distribution of the background in the condensed phase $C/F$ is more complicated than the expected fluid measure $\nu_{c_1 ,\infty}$. Together with Corollary \ref{corr} we can show that the non-zero specific relative entropy is only due to the contribution of the single condensate site and indeed the background distribution is as expected. In the following we attach some (arbitrary) ordering to the lattice sites and identify $\Lambda_L =\{ 1,\ldots ,L\}$. On $X_{L-1}$ we define the measure $\hat\pi_{L,N}$ as a marginal on the first $L-1$ coordinates
\bea\label{hatdef}
\hat\pi_{L,N} :=\pi_{L,N} (.|L\in\mathrm{argmax})^{1,..,L{-}1}
\eea
where $\pi_{L,N} (.|L\in\mathrm{argmax})$ denotes the measure $\pi_{L,N}$ conditioned on the event that $\eta_L \geq\eta_x$ for all $x=1,\ldots ,L-1$. Since $\pi_{L,N}$ is invariant under site permutations, we have
\bea\label{haty}
\hat\pi_{L,N} :=\pi_{L,N} (.|y\in\mathrm{argmax})^{\Lambda_L \setminus\{ y\}}\quad\mbox{for all }y\in\Lambda_L \ .
\eea
Note that $\hat\pi_{L,N}$ concentrates on a subset of $X_{L-1}$,
\bea\label{hatsupp}
\hat X_{L-1} :=\big\{\hat\feta\in X_{L-1} \,\big|\,\Sigma_{L-1} (\hat\feta )<N,\,\hat\eta_1 ,..,\hat\eta_{L-1} \leq N-\Sigma_{L-1} (\hat\feta )\big\}\ ,
\eea
and in case of condensation it can be interpreted as the distribution of the background.

\begin{theorem}\label{theo3}
For $\rho\geq\rho_{trans}$ we have as $L\to\infty$, $N/L\to\rho$,
\bea\label{conv3}
H(\hat\pi_{L,N} ,\nu_{c_1 ,\infty}^{L-1} )\to 0\ ,
\eea
and thus for bounded cylinder test functions
\bea\label{conv3b}
\langle f\rangle_{\pi_{L,N}}\to \langle f\rangle_{\nu_{c_1 ,\infty}}\ .
\eea
\end{theorem}

\noindent Note that the first statement (\ref{conv3}) involves the total rather than the specific relative entropy and is therefore much stronger than Corollary \ref{theo2a} and Theorem \ref{theo2}. This implies convergence in total variation norm \cite{csiszar75}. Such a result is not possible below criticality, since the conditioning on the particle number in the canonical measures leads to divergence of the relative entropy. Above criticality, this condition is accounted for purely by the condensate site and does not affect the background, which shows the same fluctuations as i.i.d.\ random variables. Following recent results in \cite{loulakisetal07}, this enables to show that the stationary density profiles converge to a Brownian motion with a jump at the location of the condensate. Below criticality the corresponding expected behaviour would be a Brownian bridge, but there is no proof so far.\\
The second statement (\ref{conv3b}) is a direct consequence of the first but not a very strong one, since it would also follow from convergence in specific relative entropy. The site with maximum occupation number will be in the support of the cylinder test function only with probability of order $1/L$. But due to this possibility, the test function has to be bounded, not necessarily by a constant but by a number of order $o(L)$. This excludes $f(\feta )=\eta_x$ as expected, since the expected density does not converge for $\rho\geq\rho_{trans}$. Note also that with (\ref{condlim}) and (\ref{conv3b}) this system is an example where weak convergence is strictly weaker than convergence in specific relative entropy.\\
\\
\textbf{Proof.} (\ref{hatdef}) and (\ref{hatsupp}) imply that
\bea
\hat\pi_{L,N} (\hat\feta )=\frac{\pi_{L,N} \big(\hat\feta ,N-\Sigma_{L-1} (\hat\feta)\big)}{\pi_{L,N} (L\in\mathrm{argmax})}\,\1_{\hat X_{L-1}} (\hat\feta )\ ,
\eea
where $(\hat\feta ,N-\Sigma_{L-1} (\hat\feta))\in X_{L,N}$ denotes the concatenated configuration. By permutation invariance we get
\bea
\pi_{L,N} (L\in\mathrm{argmax})=\frac1L\pi_{L,N} (X_{L,N}^1 ) +\tilde R_{L,N} =\frac1L\big( 1+o(1)\big)
\eea
where $0\leq\tilde R_{L,N} \leq \pi_{L,N} (X_{L,N} \setminus X_{L,N}^1 )$. So the error is exponentially small in the system size for $\rho >\rho_{trans}$ (see Corollary \ref{corr}) and of order $L^{-3/2}$ for $\rho =\rho_{trans}$.\\
Now we can compute the relative entropy
\bea\label{zwerg2}
\lefteqn{H(\hat\pi_{L,N} ,\nu_{c_1 ,\infty}^{L-1} )=\sum_{\hat\feta\in\hat X_{L-1}} \hat\pi_{L,N} (\hat\feta )\log \frac{\pi_{L,N} \big(\hat\feta ,N-\Sigma_{L-1} (\hat\feta )\big)}{\pi_{L,N} (L\in\mathrm{argmax}) \,\nu_{c_1 ,\infty}^{L-1} (\hat\feta )} =}\nonumber\\
& &=\sum_{\hat\feta\in\hat X^0_{L-1}} \!\!\!\!\hat\pi_{L,N} (\hat\feta )\log\frac{w_R^{L-1} (\hat\feta ) c_0^{-R} c_1^{-(N-\Sigma_{L-1} (\hat\feta )-R)} L z_\infty^{L-1} (c_1 )}{(c_1 /c_0)^{\Sigma_{L-1} (\hat\feta)}\, Z_{L,N}\, (1+o(1))} {+}R_{L,N} \, ,
\eea
where analogous to (\ref{decompose})
\bea
\hat X^0_{L-1} =\big\{\hat\feta\in \hat X_{L-1}\,\big|\,\hat\eta_1 ,\ldots ,\hat\eta_{L-1} \leq R\big\}\ .
\eea
Therefore we have
\bea
|R_{L,N} |&=&\bigg|\sum_{\hat\feta\in\hat X_{L-1} \setminus \hat X^0_{L-1}} \hat\pi_{L,N} (\hat\feta )\log \frac{\pi_{L,N} \big(\hat\feta ,N-\Sigma_{L-1} (\hat\feta )\big)\, L}{\nu_{c_1 ,\infty}^{L-1} (\hat\feta )\, (1+o(1))}\bigg|\leq\nonumber\\
&\leq &C\pi_{L,N} \big( X_{L,N} \setminus (X_{L,N}^0 \cup X_{L,N}^1 )\big)\, L\to 0\quad\mbox{as }L\to\infty\ ,
\eea
using Corollary \ref{corr} and Theorem \ref{theo1b}, since the argument of the logarithm is at most exponential in $L$. On $\hat X_{L-1}^0$ we have $w_R^{L-1} (\hat\feta )=c_0^{-\Sigma_{L-1} (\hat\feta )}$ and thus
\bea\label{hzwerg}
H(\hat\pi_{L,N} ,\nu_{c_1 ,\infty}^{L-1} )=\pi_{L,N} (X_{L,N}^1 )\log\frac{(c_1 /c_0 )^R \, c_1^{-N} \, L\, c_0^{L-1}}{Z_{L,N} \, (c_0 -c_1)^{L-1}} +o(1)\ ,
\eea
where we have used $\hat\pi_{L,N} (\hat X_{L-1}^0 )=\pi_{L,N} (X_{L,N}^1 )$. With Theorems \ref{theo1} and \ref{theo1b} we have for $\rho\geq\rho_{trans}$
\bea
Z_{L,N} &=&w_R^L (X_{L,N}^1) \big( 1+o(1)\big) =\nonumber\\
&=&L^{3/2} \Big(\frac{c_1}{c_0}\Big)^{R+\rho_c L} c_1^{-N} \big| X_{L-1,\rho_c (L-1)}^0 \big|\,\sqrt{2\pi \rho_c (1+\rho_c )}\big(1+o(1)\big)\ ,
\eea
and according to Lemma \ref{lemma1}
\bea
\big| X_{L-1,\rho_c (L-1)}^0 \big| ={(1+\rho_c )(L-1)-1\choose L-2 } \big( 1+o(1)\big)\ .
\eea
A careful application of Stirling's formula, which was not necessary in the proof of Theorem \ref{theo1b}, yields
\bea
{(1{+}\rho_c )(L{-}1){-}1\choose L-2 } {=}\Big(\frac{c_0}{c_0 {-}c_1}\, c_0^{\rho_c}\Big)^{L-1} \big( 2\pi\rho_c (1{+}\rho_c )L\big)^{-1/2} \big( 1{+}o(1)\big)\, ,
\eea
where we have used in the exponential term that $\rho_c =c_1 /(c_0 -c_1 )$. Plugging everything into (\ref{hzwerg}) this leads to a perfect cancellation and we get as $L\to\infty$
\bea
H(\hat\pi_{L,N} ,\nu_{c_1 ,\infty}^{L-1} )=\pi_{L,N} (X_{L,N}^1 )\log\big( 1+o(1)\big) +o(1)\to 0\ ,
\eea
which finishes the proof of the first statement.

\noindent Let $f\in C_{0,b} (X)$ be a cylinder test function bounded by $C$ and supported on the lattice sites $supp (f)\subset\N$ with $\big| supp(f)\big| =n$. In the following let $L>\max supp(f)$ such that $supp (f)\subsetneq\Lambda_L$. We have
\bea
\langle f\rangle_{\pi_{L,N}} =\sum_{\feta\in X_{L,N}} \pi_{L,N} (\feta )\, f(\feta )=\sum_{\feta\in X^1_{L,N}} \pi_{L,N} (\feta )\, f(\feta )+R_{L,N}^1 \ ,
\eea
where due to Corollary \ref{corr}
\bea
|R_{L,N}^1 |=\Big|\sum_{\feta\in X_{L,N}\setminus X_{L,N}^1} \pi_{L,N} (\feta )\, f(\feta )\Big|\leq C\pi_{L,N} (X_{L,N}\setminus X_{L,N}^1 )\to 0\ ,
\eea
Since $|\mathrm{argmax}(\feta )|=1$ for all $\feta\in X_{L,N}^1$, we have
\bea
\lefteqn{\sum_{\feta\in X^1_{L,N}} \pi_{L,N} (\feta )\, f(\feta )=\frac1L \sum_{y\in\Lambda_L} \sum_{\feta\in X^1_{L,N}} \pi_{L,N} \big(\feta\,\big|\,\mathrm{argmax} =\{ y\}\big)\, f(\feta )=}\nonumber\\
& &=\frac1L \sum_{y\in\Lambda_L \setminus supp(f)} \sum_{\hat\feta\in\hat X^0_{L-1}} \pi_{L,N} \big(\hat\feta\,\big|\,\mathrm{argmax} =\{ y\}\big)^{\Lambda_L \setminus\{ y\}}\, f(\hat\feta )+R_{L,N}^2 \nonumber\\
& &=\frac{L-n}{L}\sum_{\hat\feta\in\hat X^0_{L-1}} \hat\pi_{L,N} (\hat\feta)\, f(\hat\feta )+R_{L,N}^2
\eea
due to (\ref{haty}), where boundedness of $f$ implies
\bea
|R_{L,N}^2 |&=&\bigg|\frac1L \sum_{y\in supp(f)} \sum_{\feta\in X^1_{L,N}} \pi_{L,N} \big(\feta\,\big|\,\mathrm{argmax} =\{ y\}\big)\, f(\hat\feta )\bigg|\nonumber\\
&\leq &\frac{n}{L}\, C\,\pi_{L,N} (X_{L,N}^1 )\to 0\quad\mbox{as }L\to\infty\ .
\eea
Note that this is the only place where we crucially require that $f$ is bounded by a constant of order $o(L)$. With Corollary \ref{corr} we get
\bea
\sum_{\hat\feta\in\hat X^0_{L-1}} \hat\pi_{L,N} (\hat\feta)\, f(\hat\feta )=\langle f\rangle_{\hat\pi_{L,N}} +R_{L,N}^3
\eea
where
\bea
|R_{L,N}^3 |&=&\bigg|\sum_{\hat\feta\in\hat X_{L-1}\setminus\hat X^0_{L-1}} \hat\pi_{L,N} (\hat\feta)\, f(\hat\feta )\bigg|\leq C\,\hat\pi_{L,N} (\hat X_{L-1}\setminus\hat X^0_{L-1})=\nonumber\\
&=&C\,\pi_{L,N} \big( X_{L,N} \setminus (X_{L,N}^0 \cup X_{L,N}^1 )\big)\to 0\quad\mbox{as }L\to\infty\ .
\eea
Together with (\ref{conv3}) shown above, this implies 
\bea
\langle f\rangle_{\pi_{L,N}} =\langle f\rangle_{\hat \pi_{L,N}} +R_{L,N}^1 +R_{L,N}^2 +R_{L,N}^3 \to\langle f\rangle_{\nu_{c_1 ,\infty}} \ ,
\eea
since convergence in total relative entropy implies weak convergence.\hfill $\Box$

\section{Metastability}

In the previous section the role of the density $\rho_c +a$ remains open. A first hint appears in the proof of Theorem \ref{theo1}, where the saddle point equation for condensate contributions (\ref{saddle}) only has solutions for $\rho\geq\rho_c +a$. But in the context of the equivalence of ensembles we are not able to distinguish the phases $F$ and $F/C$ in the phase diagram (Figure \ref{fig1}), as can be seen in Table \ref{table1}. A further analysis of the canonical measures 
in terms of the order parameter of the model, i.e. the background density $\rho_{bg}$ of uncondensed particles, will clarify this point. Since for $a=0$ the condensation transition is continuous, we only consider the case $a>0$ throughout this section. We define the observable
\bea\label{sigmabg}
\Sigma_L^{bg} (\feta ):=\Sigma_L (\feta )-\max_{x\in\Lambda_L} \eta_x \ ,
\eea
which can be interpreted as the number of particles in the background, 
since at most one site contributes to the condensate.

\begin{theorem}\label{theo4}
Let $S_1 ,S_2 ,\ldots\in\N$ be any sequence with $S_L /L\to\rho_{bg} >0$. Then the limit
\bea\label{ratef}
I_\rho (\rho_{bg} ):=-\lim_{L\to\infty} \frac1L\log \pi_{L,N} \big(\Sigma_L^{bg} =S_L \big)\in [0,\infty ]
\eea
exists for all $\rho >0$ ($N/L\to\rho$), and defines the rate function for the events $\big\{\Sigma_L^{bg} =S_L \big\}$. For $\rho_{bg} >\rho$, $I_\rho (\rho_{bg})=\infty$ and for $\rho_{bg} \leq\rho$ it can be written as
\bea\label{ratefct}
I_\rho (\rho_{bg} )&=&s_{can} (\rho ) -s_{fluid} (\rho_{bg}) +\nonumber\\
& &+\left\{\bacl (\rho -\rho_{bg} )\log c_0 \ &,\ \rho_{bg} \geq\rho -a\\ (\rho -\rho_{bg} )\log c_1 +a\log (c_0 /c_1 )\ &,\ \rho_{bg} \leq\rho -a\ea\right.\ .
\eea
\end{theorem}

\noindent\textbf{Proof.} For $\rho_{bg} >\rho$, $S_L >N$ eventually and thus $\pi_{L,N} \big(\Sigma_L^{bg} =S_L \big) =0$ eventually. For $\rho_{bg} \leq\rho$ we use the identity
\bea
\pi_{L,N} =\nu^L_{\phi_R (\rho ),R} (\, .\,|\,\Sigma_L =N )=\frac{\nu^L_{\phi_R (\rho ),R} \big(\, .\cup\{\Sigma_L =N\}\big)}{\nu^L_{\phi_R (\rho ),R} (\Sigma_L =N )}
\eea
and the fact that (\ref{relent2}) and (\ref{relent3}) imply
\bea
{-}\frac1L\log\nu^L_{\phi_R (\rho ),R} (\Sigma_L {=}N )=h\big(\pi_{L,N} ,\nu^L_{\phi_R (\rho ),R}\big)\to s_{gcan} (\rho )- s_{can} (\rho )\, .
\eea
Furthermore, Corollary \ref{corr} implies that
\bea\label{impl}
\lim_{L\to\infty}\frac1L\log\pi_{L,N} (.)=\lim_{L\to\infty}\frac1L\log\pi_{L,N} \big(\, .\,\cap (X_{L,N}^0 \cup X_{L,N}^1 )\big)\ ,
\eea
and therefore we get
\bea\label{drei}
\lefteqn{\lim_{L\to\infty} \frac1L\log \pi_{L,N} \big(\Sigma_L^{bg} = S_L \big) =s_{gcan} (\rho )-s_{can} (\rho ) +}\nonumber\\
& &+\lim_{L\to\infty} \frac1L\log\nu_{\phi_R (\rho ),R} (\eta_L =N-S_L )+\nonumber\\
& &+\lim_{L\to\infty} \frac1L\log\nu_{\phi_R (\rho ),R}^{L-1} \big(\Sigma_{L-1} =S_L ,\eta_1 ,..,\eta_{L-1} \leq (N-S_L )\wedge R\big)\ .
\eea
For the last two terms we have fixed the maximum to be on site $L$, since the corresponding polynomial correction vanishes on the logarithmic scale in the limit. With the definition of the single site measure (\ref{mass}) the second last term is given by
\bea\label{eins}
\lefteqn{\lim_{L\to\infty} \frac1L\log\nu_{\phi_R (\rho ),R} (\eta_L =N-S_L )=}\nonumber\\
& &\quad =\left\{\bacl (\rho -\rho_{bg} )\log (\phi_{gcan} (\rho )/c_0 )\ &,\ \rho_{bg} \geq\rho -a\\ (\rho -\rho_{bg} )\log (\phi_{gcan} (\rho )/c_1 )-a\log (c_0 /c_1 )\ &,\ \rho_{bg} \leq\rho -a\ea\right.\ ,
\eea
where (cf. (\ref{phir}))
\bea
\phi_{gcan} (\rho ):=\lim_{L\to\infty} \phi_R (\rho )=\left\{\bacl \phi_\infty (\rho )=c_0 \rho /(1+\rho)\ &,\ \rho\leq\rho_c \\ c_1 \ &,\ \rho\geq\rho_c \ea\right.\ .
\eea
Due to the condition $\eta_1 ,\ldots ,\eta_{L-1} \leq (N-S_L )\wedge R$ in the last term, which follows from (\ref{impl}) and (\ref{sigmabg}), all configurations in that event have the same probability and we get
\bea\label{zwei}
\lefteqn{\lim_{L\to\infty} \frac1L\log\nu_{\phi_R (\rho ),R}^{L-1} \big(\Sigma_{L-1} =S_L ,\eta_1 ,..,\eta_{L-1} \leq (N-S_L )\wedge R\big) =}\nonumber\\
& &\qquad =\lim_{L\to\infty} \frac1L\log\bigg(\frac{(\phi_R (\rho )/c_0 )^{S_L}}{z_R (\phi_R (\rho ))^{L-1}} |\tilde X_{L-1,S_L}^0 |\bigg) =\nonumber\\
& &\qquad =\rho_{bg}\log\frac{\phi_{gcan} (\rho )}{c_0}-p\big(\phi_{gcan}(\rho)\big) +\chi (\rho_{bg} )=\nonumber\\
& &\qquad =(\rho_{bg} -\rho )\log\phi_{gcan} (\rho )-s_{gcan} (\rho ) +s_{fluid} (\rho_{bg})\ .
\eea
where
\bea
\tilde X_{L-1,S_L}^0 =\big\{\feta\in X_{L-1 ,S_L} \,\big|\, \eta_1 ,\ldots ,\eta_{L-1} \leq (N-S_L )\wedge R\big\} \ .
\eea
Due to the more restrictive condition this is only a subset of $X_{L-1,S_L}^0$, but completely analogously to Lemma \ref{lemma1} one can show that as $L\to\infty$
\bea
\frac1L\log |\tilde X_{L-1,S_L}^0 |\to (1{+}\rho_{bg} )\log (1{+}\rho_{bg} )- \rho_{bg}\log\rho_{bg} =\chi (\rho_{bg} )\ .
\eea
Inserting (\ref{eins}) and (\ref{zwei}) into (\ref{drei}) finishes the proof.\hfill $\Box$\\

\begin{figure}
\begin{center}
\includegraphics[width=0.48\textwidth]{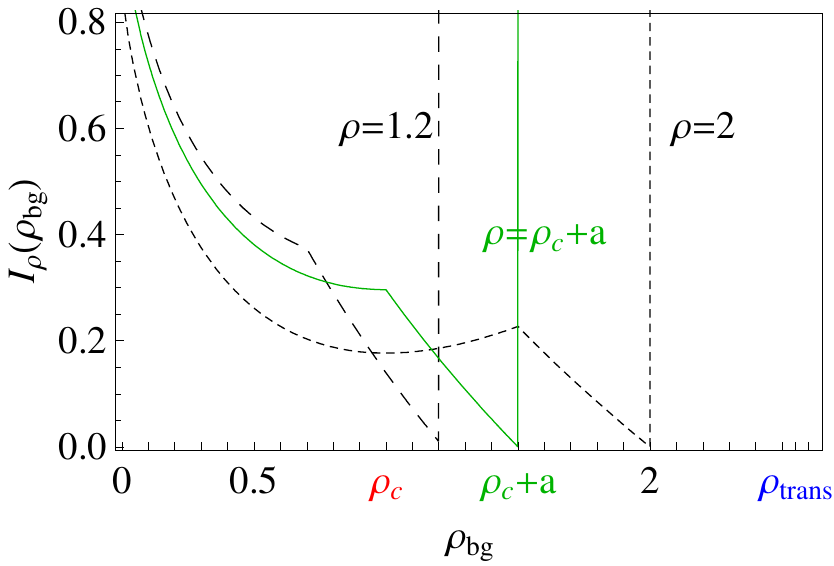}\hfill
\includegraphics[width=0.48\textwidth]{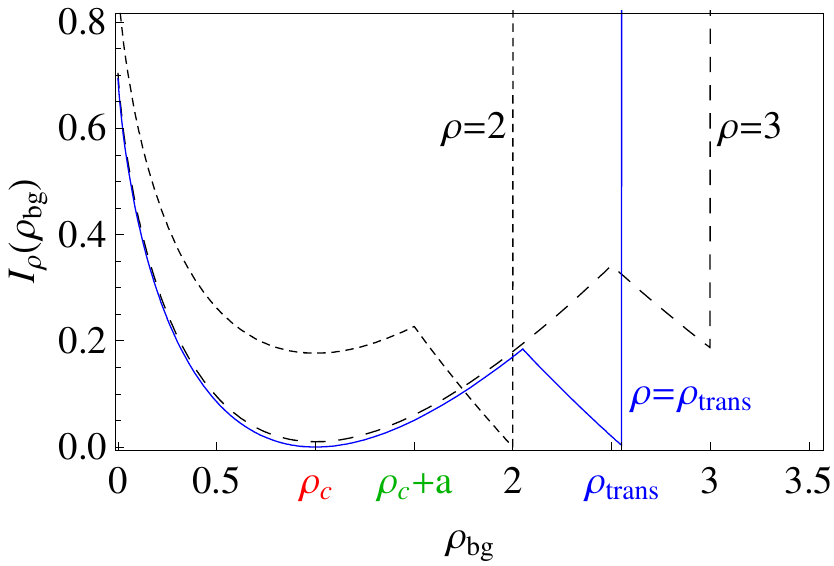}
\end{center}
\caption{\label{fig4}
The rate function $I_\rho (\rho_{bg})$ for $c_0 =2$, $c_1 =1$, $a=0.5$ and various values of $\rho$. For $\rho >\rho_c +a$ the function has a local minimum at $\rho_{bg} =\rho_c$, which becomes the global minimum for $\rho >\rho_{trans}$.}
\end{figure}

\noindent Figure \ref{fig4} shows that the distribution of $\Sigma_L^{bg}$ concentrates on values of the order $\rho L$ for $\rho <\rho_{trans}$ and on values of the order $\rho_c L$ for $\rho >\rho_{trans}$. These two cases correspond to the phases $F/C$ and $C/F$, respectively, and have been identified already in the previous section. But in Figure \ref{fig4} also the role of $\rho_c +a$ can be identified. For $\rho <\rho_c +a$, the rate function $I_\rho (\rho_{bg} )$ (\ref{ratef}) has only one minimum $I_\rho (\rho )=0$, whereas for $\rho >\rho_c +a$ it has an additional local minimum at $\rho_{bg} =\rho_c$, i.e. the condensed phase becomes metastable. For $\rho >\rho_{trans}$ this local minimum becomes the global one, and the fluid phase becomes metastable. For $\rho =\rho_{trans}$ the rate function vanishes for both phases, but the finer analysis of Theorem \ref{theo1b} reveals that the fluid phase is already metastable in this case.

By definition, the observable $\Sigma_L^{bg} (\feta )$ changes at most by $\pm 1$ during each jump of a particle. So the process $\big(\Sigma_L^{bg} (\feta (t))\big)_{t\geq 0}$ is a one-dimensional simple random walk (or a birth-death process) on $\{ 0,1,\ldots ,N\}$, whose stationary large deviation rate function is $I_\rho$. The minima of this rate function correspond to the fluid phase for $\rho_{bg} =\rho$ and the condensed phase for $\rho_{bg} =\rho_c$. For finite $L$ the system has two quasi-stationary distributions
\bea
\pi_{L,N} (.|X_{L,N}^0 )\quad\mbox{and}\quad \pi_{L,N} (.|X_{L,N}^1 )\ ,
\eea
corresponding to the fluid and the condensed phase, respectively. Analogous to (\ref{hatdef}) we define
\bea
\tilde \pi_{L,N} =\pi_{L,N} (.|X_{L,N}^1 ,\, L\in\mathrm{argmax})^{1,..,L-1}
\eea
%\pagebreak
\begin{proposition}\label{prop3}
In the limit $L\to\infty$, $N/L\to\rho$, we have for all $\rho\geq 0$
\bea\label{prop31}
h\big(\pi_{L,N} (.|X_{L,N}^0 ),\nu^L_{\phi_\infty (\rho ),\infty}\big)\to 0\quad\mbox{and}\quad \pi_{L,N} (.|X_{L,N}^0 )\to \nu_{\phi_\infty (\rho ),\infty}\ ,
\eea
and for all $\rho\geq \rho_c +a$
\bea\label{prop32}
H\big(\tilde\pi_{L,N},\nu^{L-1}_{c_1 ,\infty}\big)\to 0\quad\mbox{and}\quad\pi_{L,N} (.|X_{L,N}^1 )\to \nu_{c_1 ,\infty}\ .
\eea
In both cases the second convergence is weakly with respect to bounded cylinder test functions.
\end{proposition}

\noindent As in Theorem \ref{theo3}, we can show convergence in total relative entropy (\ref{prop32}) for the condensed phase, which is much stronger than convergence in specific relative entropy (see comments in the previous section).\\

\noindent\textbf{Proof.} The first statements in (\ref{prop31}) and (\ref{prop32}) can be proved analogous to Theorem \ref{theo2} and Theorem \ref{theo3}, respectively. Since
\bea
\pi_{L,N} (\feta |X_{L,N}^0 )=\frac{\pi_{L,N} (\feta )}{\pi_{L,N} (X_{L,N}^0 )} \1_{X_{L,N}^0} (\feta )=\frac{1}{|X_{L,N}^0 |} \1_{X_{L,N}^0} (\feta )
\eea
is the uniform measure on $X_{L,N}^0$, we get for (\ref{prop31}) analogous to (\ref{zwerg})
\bea
\lefteqn{h\big(\pi_{L,N} (.|X_{L,N}^0 ),\nu_{\phi_\infty (\rho ),\infty}^L \big) =}\nonumber\\
& &\qquad =\frac1L \sum_{\feta\in X_{L,N}^0} \pi_{L,N} (\feta |X_{L,N}^0)\log \frac{z_\infty \big(\phi_\infty (\rho )\big)^L}{\big(\phi_\infty (\rho )/c_0 \big)^N |X_{L,N}^0 |} =\nonumber\\
& &\qquad= p_{fluid} \big(\phi_\infty (\rho )\big) -\frac{N}{L}\,\log\frac{\phi_\infty (\rho )}{c_0} -\frac1L\,\log |X_{L,N}^0 |\nonumber\\
& &\qquad\to s_{fluid} (\rho )+\rho\log c_0 -\chi (\rho )=0\quad\mbox{as }L\to\infty
\eea
for all $\rho\geq 0$, using Lemma \ref{lemma1}. Note that here we are a priori restricted to $X_{L,N}^0$ so that there is no error term as in (\ref{zwerg}).\\
The same holds for a modification of (\ref{zwerg2}) to derive (\ref{prop32}). Using
\bea
\tilde\pi_{L,N} (\hat\feta )&=&\frac{\pi_{L,N} (\hat\feta ,N-\Sigma_{L-1} (\hat\feta ))}{\pi_{L,N} (\mathrm{argmax}=L,X_{L,N}^1 )}\,\1_{\mathrm{argmax}=L,X_{L,N}^1}\big(\hat\feta ,N{-}\Sigma_{L{-}1} (\hat\feta )\big) =\nonumber\\
&=&\frac{L (c_0 /c_1 )^{{-}\Sigma_{L-1} (\hat\feta )-R} c_1^{-N}}{\pi_{L,N} (X_{L,N}^1 )\, Z_{L,N}}\1_{\mathrm{argmax}=L,X_{L,N}^1}\big(\hat\feta ,N{-}\Sigma_{L{-}1} (\hat\feta )\big)
\eea
and $\pi_{L,N} (X_{L,N}^1 )\to 1$, we get in direct analogy to the proof of Theorem \ref{theo3}
\bea\label{doedle}
H(\tilde\pi_{L,N} ,\nu_{c_1 ,\infty}^{L-1} )&=&\sum_{\hat\feta\in\hat X_{L-1}^0} \tilde\pi_{L,N} (\hat\feta )\log \frac{L\, (c_0 /c_1 )^{-R} \, c_1^{-N} z_\infty (c_1 )^{L{-}1}}{\pi_{L,N} (X_{L,N}^1 )\, Z_{L,N}} =\nonumber\\
&=&\pi_{L,N} (X_{L,N}^1 )\log\big( 1+o(1)\big)\to 0\quad\mbox{as }L\to\infty\ .
\eea

The second statements in (\ref{prop31}) and (\ref{prop32}) follow completely analogously to the proofs of Theorem \ref{theo2} and Theorem \ref{theo3}.\hfill $\Box$\\
\\
With Theorems \ref{theo2} and \ref{theo3} both statements of Proposition \ref{prop3} follow directly from Corollary \ref{corr} and Theorem \ref{theo1b}, but only for $\rho <\rho_{trans}$ and $\rho \geq\rho_{trans}$, respectively, where the quasi-stationary distributions converge to the stationary distribution.

For finite $L$, both phases have life-times of the order $\sim e^{\xi (\rho )L}$ exponential in the system size for all $\rho >\rho_c +a$, where the exponential rate $\xi (\rho )$ depends on the density. It can be calculated using the hitting times
\bea
\tau_L^{fluid} (\rho )&:=&\inf\big\{ t\geq 0\,\big|\,\max_{x\in\Lambda_L} \eta_x (t)>R\big\}\ ,\nonumber\\
\tau_L^{cond} (\rho )&:=&\inf\big\{ t\geq 0\,\big|\,\max_{x\in\Lambda_L} \eta_x (t)\leq R\big\}\ ,
\eea
which depend on the initial configuration as well as the time evolution. Due to the effective one-dimensional random walk picture mentioned above, the quasi-stationary expectations of these random variables are determined by the rate functions at the locations of local minima and maxima. These are
\bea
I_\rho (\rho )&=&s_{can} (\rho )-s_{fluid} (\rho )\qquad\qquad\qquad\qquad\qquad\qquad\ \mbox{(min.)}\nonumber\\
I_\rho (\rho -a)&=&s_{can} (\rho )-s_{fluid} (\rho -a)+a\log c_0 \qquad\qquad\qquad\ \mbox{(max.)}\nonumber\\
I_\rho (\rho_c )&=&s_{can} (\rho )-s_{fluid} (\rho_c )+a\log\frac{c_0}{c_1} +(\rho -\rho_c )\log c_1\,\,\mbox{(min.)}\, ,
\eea
where the last two are only defined for $\rho >\rho_c +a$ (cf. Figure \ref{fig4}). Note that $I_\rho (\rho )=0$ for $\rho \leq\rho_c$, whereas $I_\rho (\rho_c )=0$ for $\rho \geq\rho_c$. For $\rho >\rho_c +a$ we then have
\bea\label{lifetimes}
\xi^{fluid} (\rho )&:=&\lim_{L\to\infty}\frac1L\log\ll \tau_L^{fluid} (\rho )\gg_{\pi_{L,N} (.|X_{L,N}^0), e^{\mathcal{L}t}} =I_\rho (\rho -a)-I_\rho (\rho )=\nonumber\\
&=& s_{fluid} (\rho )-s_{fluid} (\rho -a)+a\log c_0\ ,\nonumber\\
\xi^{cond} (\rho )&:=&\lim_{L\to\infty}\frac1L\log\ll \tau_L^{cond} (\rho )\gg_{\pi_{L,N} (.|X_{L,N}^1), e^{\mathcal{L}t}} =I_\rho (\rho -a)-I_\rho (\rho_c )=\nonumber\\
&=& s_{fluid} (\rho_c )-s_{fluid} (\rho -a)+(\rho_c +a-\rho )\log c_1
\eea
where $\ll ..\gg_{\pi_{L,N} (.|X_{L,N}^0), e^{\mathcal{L}t}}$ denotes the average with respect to a quasi-stationary initial distribution and the time evolution given by the generator $\mathcal L$ (\ref{generator}). Note that for $\rho <\rho_c +a$, $\xi_{fluid} (\rho )=\infty$ and $\xi_{cond} (\rho )$ is not defined, since the condensed phase is not stable. The asymptotic behaviour as $\rho\to\infty$ is given by
\bea
\xi^{fluid} (\rho )&\simeq &\log\frac{1+\rho}{1+\rho -a}\to 0\nonumber\\
\xi^{cond} (\rho )&\simeq &\rho\log\frac{c_0}{c_1}-\log (1+\rho -a)+const.\to\infty\ .
\eea

\begin{figure}
\begin{center}
\includegraphics[width=0.5\textwidth]{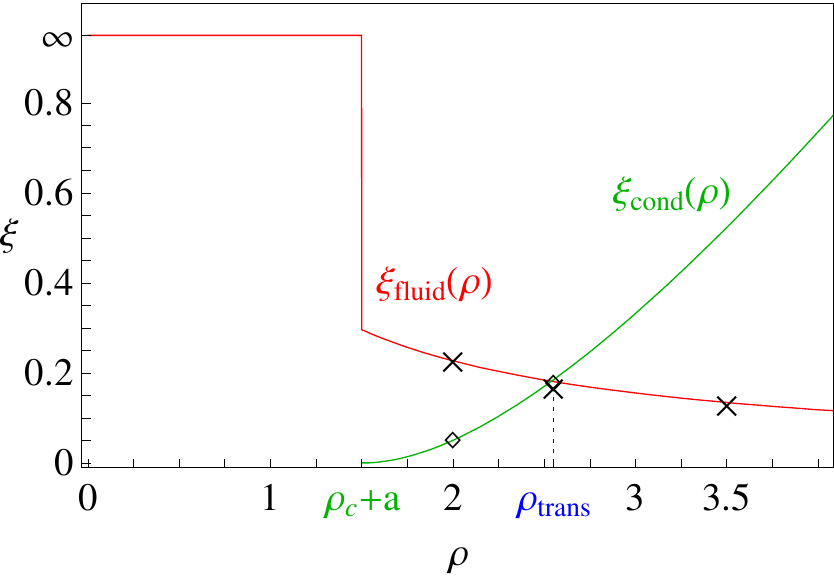}\\[5mm]
\includegraphics[width=0.48\textwidth]{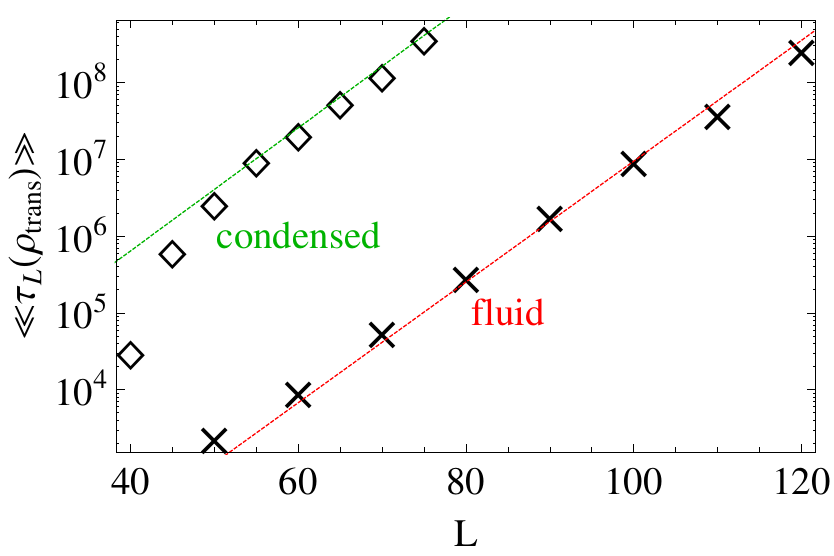}\quad
\includegraphics[width=0.48\textwidth]{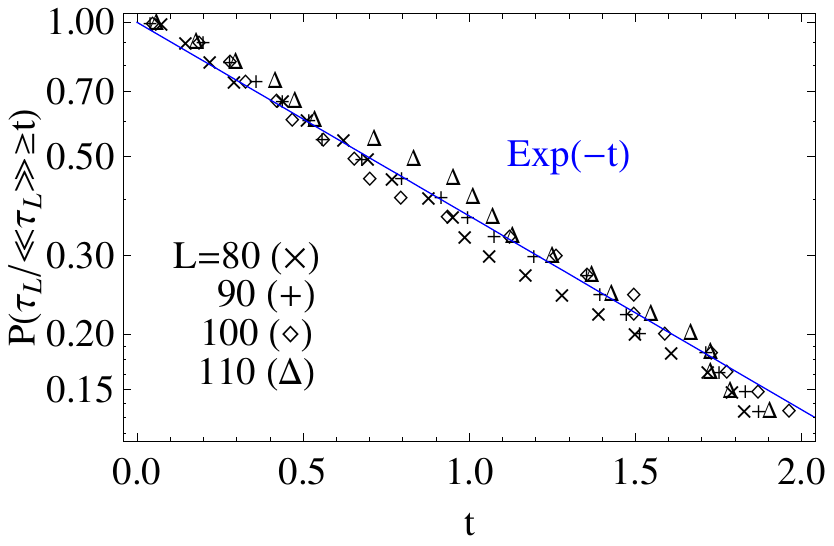}
\end{center}
\caption{\label{fig5}
Life-times of fluid and condensed phase for $c_0 =2$, $c_1 =1$, $a=0.5$.\newline
Top: The exponential rate $\xi (\rho )$ of the life-time as a function of the density. $\times$ and $\diamond$ denote Monte Carlo data, errors are of the size of the symbols. Bottom left: Expected life-times as used in (\ref{lifetimes}) in a logarithmic plot as a function of $L$ for $\rho =\rho_{trans}$. Bottom right: Tail distribution of the normalized lifetimes $\tau_L^{cond} (\rho_{trans}) /\ll \tau_L^{cond} (\rho_{trans})\gg$, compared with the tail of an $\mathit{Exp}(1)$ random variable.}
\end{figure}

\noindent For all $a>0$ we have $\xi^{fluid} (\rho_c +a)>0=\xi^{cond} (\rho_c +a)$ and $\xi^{fluid} (\rho_{trans} )=\xi^{cond} (\rho_{trans} )$, as expected. This behaviour is illustrated in Figure \ref{fig5} for some specific values of the parameters. The predictions are in very good agreement with data from Monte Carlo simulations, a few of which are presented in the figure. On the bottom left for $\rho =\rho_{trans}$ we see that the expected lifetimes for the condensed phase are larger than for the fluid phase, which is in accordance with Theorem \ref{theo1b}. There appears to be a polynomial correction in the condensed phase, but the data are not good enough to measure the power in $L$.

Note that the last part of the derivation in this section is not rigorous, since strictly speaking $\big(\Sigma_L^{bg} (\feta (t))\big)_{t\geq 0}$ is not a Markov process. Still one could use a potential theoretic approach analogous to \cite{bovier}, to show rigorously that the average life times of both phases are exponential in $L$. However, getting the right timescale with this approach would require quite some technical effort. Besides the exponential growth rate of the life times with the system size $L$, simulations also indicate that the distribution of the lifetimes is actually exponential, as can be seen in Figure \ref{fig5} on the bottom right. This is to be expected, since the system effectively jumps between the two metastable phases in a Markovian way.

\section{Dependence on the number of particles}

In this section we consider the case where the jump rates depend on the number of particles in the system rather than the lattice size, which is also the case in some models for granular clustering \cite{lipowskietal02,coppexetal02}, one of our main motivations for this study. We modify our original model (\ref{rates}),
\bea\label{rates3}
g_R (k)=\left\{\bacl c_0 \ &,\ k\leq R \\ c_1 \ &,\ k> R\ea\right.\quad\mbox{for }k\geq 1\ ,\quad g(0)=0\ ,
\eea
where $R$ is now a function of the number of particles $\Sigma_L (\feta )$. For simplicity we concentrate on the specific choice $R=a\,\Sigma_L (\feta )$ with $a\in [0,1)$, since $a\geq 1$ is not interesting for this model. So in principle, the jump rates do not only depend on the local occupation number but on the global configuration. But restricted to a subset $X_{L,N}$ with fixed particle number $\Sigma_L (\feta ) =N$, $R$ is just a parameter, the process is well defined and standard results on stationary measures apply. Therefore the canonical measures are well defined as in (\ref{canens}). In particular, Theorems \ref{theo1} to \ref{theo3} still hold and the proofs apply directly, where $a$ should be replaced by $a\rho$, since now $R/L\to\rho\, a$. So analogous to (\ref{acon}) the transition density $\rho_{trans}$ is determined by the relation
\bea\label{acon2}
a=\Big( s_{fluid} (\rho_c )-(\rho -\rho_c )\log c_1 -s_{fluid} (\rho )\Big)\Big/\Big(\rho\log\frac{c_0}{c_1}\Big) \ ,
\eea
and $s_{fluid}$ is given as in (\ref{sfluid}). The canonical entropy density $s_{can}$ is still given by (\ref{scan2}), but the contribution of the condensate, which determines the behaviour for large $\rho$, is now given by
\bea\label{scond2}
s_{cond} (\rho ,\rho_c )=-\rho \Big( a\log\frac{c_0}{c_1}+\log c_1 \Big) +\rho_c \log c_1 \ .
\eea
This leads to
\bea\label{scann}
s_{can} (\rho )=\left\{\bacl s_{fluid} (\rho )\ &,\ \rho \leq\rho_{trans}\\ s_{fluid} (\rho_c ){-}\rho \Big( a\log\frac{c_0}{c_1}{+}\log c_1 \Big) {+}\rho_c \log c_1\ &,\ \rho >\rho_{trans}\ea\right. .
\eea

To study the equivalence of ensembles, one has to define the grand-canonical measures. This is not as straightforward as in (\ref{mass}), since the number of particles $\Sigma_L (\feta )$ and thus $R$ is now a random variable. However, we know that the set of all stationary measures is convex, and the extremal points are the canonical measures (see e.g. \cite{liggett} or \cite{stefan2}). So the grand-canonical measures can be defined as convex combinations of canonical measures,
\bea
\nu^L_{\phi ,R} (\feta )=\prod_{x\in\Lambda_L} w_R^L (\eta_x )\,\phi^{\eta_x} \Big/\sum_{N=0}^\infty \phi^N Z_{L,N}\ .
\eea
If the weights $w_R (k)$ depended only on the system size $L$, this would be equivalent to (\ref{mass}), but here the measures are obviously not of product form since the weights depend on the total number of particles through $R=a\Sigma_L (\feta )$. Also the normalizing partition function
\bea
Z_R (\phi ):=\sum_{N=0}^\infty \phi^N Z_{L,N}
\eea
does not factorize, since now $Z_{L,N} =w_{aN}^L (X_{L,N})$. Nevertheless we can define the pressure
\bea\label{druck}
p_{gcan} (\phi ):=\lim_{L\to\infty} \frac1L\log\sum_{N=0}^\infty \phi^N Z_{L,N}\ ,
\eea
and by a saddle point argument analogous to the proof of Theorem \ref{theo1} this is well defined and given by
\bea\label{saddle2}
p_{gcan} (\phi )=\sup_{\rho\geq 0} \big(\rho\log\phi +s_{can} (\rho )\big)\ ,
\eea
the Legendre transform of the negative canonical entropy density (\ref{scann}). With (\ref{scond2}) we have for $\rho >\rho_{trans}$
\bea
\rho\log\phi +s_{can} (\rho )= \rho \Big(\log\phi - a\log\frac{c_0}{c_1}-\log c_1 \Big) +\rho_c \log c_1 \ ,
\eea
which, analogously to (\ref{pgcan}), implies
\bea\label{pgcan2}
p_{gcan} (\phi )=\left\{\bacl p_{fluid} (\phi )\ &,\ \phi <\phi_c (a)\\ \infty\ &,\ \phi \geq\phi_c (a) \ea\right.\ .
\eea

\begin{figure}
\begin{center}
\includegraphics[width=0.43\textwidth]{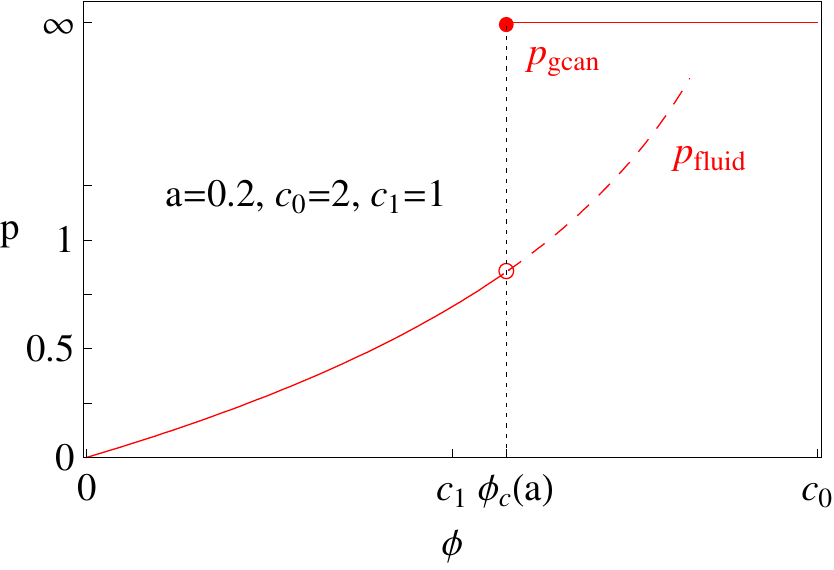}\quad\includegraphics[width=0.47\textwidth]{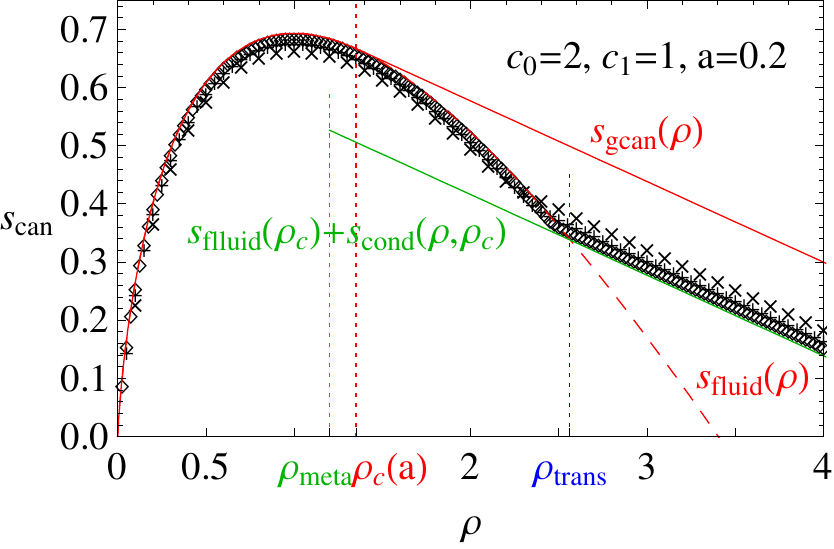}\\
\end{center}
\caption{\label{fig6a} Pressure and entropies for $a=0.2$, $c_0 =2$ and $c_1 =1$ as given in (\ref{pgcan2}), (\ref{scann}) and (\ref{sgcan2}). Data points are calculated numerically according to (\ref{recursion}) with $L=100\ (\times )$, $200\ (+)$, $400\ (\Diamond)$, and show good agreement with the theoretical predictions for the thermodynamic limit.}
\end{figure}

\noindent The difference is that now the pressure is finite up to
\bea
\phi_c (a):=c_1 \Big(\frac{c_0}{c_1}\Big)^a \geq c_1 \ ,
\eea
which is strictly bigger than the value $c_1$ in (\ref{pgcan}) for all $a\in (0,1)$. Note that for $\phi =\phi_c (a)$ the saddle point argument (\ref{saddle2}) does not apply and (\ref{druck}) diverges, so $p_{gcan} \big(\phi_c (a)\big) =\infty$, see Figure \ref{fig6a} left. As a consequence of (\ref{pgcan2}), the critical density defined as in (\ref{rhoc}) is now $a$-dependent and given by
\bea\label{rhoca}
\rho_c (a):=\rho_\infty \big(\phi_c (a)\big) =\frac{c_1^{1-a}}{c_0^{1-a} -c_1^{1-a}}\ ,
\eea
where $\rho_\infty$ is still given by (\ref{rhoinf}). So analogous to (\ref{sgcan}), the grand-canonical entropy density is given by the negative Legendre transform of (\ref{pgcan2}),
\bea\label{sgcan2}
s_{gcan} (\rho )=\left\{\bacl s_{fluid} (\rho )\ &,\ \rho\leq\rho_c (a)\\ s_{fluid} (\rho_c (a))-(\rho -\rho_c (a))\log\phi_c (a)\ &,\ \rho >\rho_c (a)\ea\right.\ .
\eea
By definition, this is again the concave hull of $s_{can} (\rho )$, as can be seen in Figure \ref{fig6a}, right. The canonical entropy density is calculated numerically using (\ref{recursion}) for different values of $L$ and $N$, and as before the results agree very well with the predictions.

\begin{figure}
\begin{center}
\includegraphics[width=0.5\textwidth]{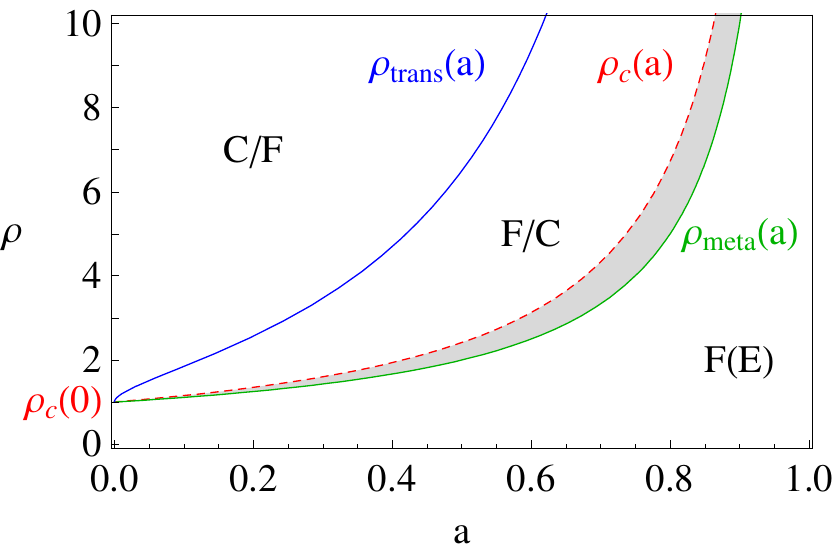}
\end{center}
\caption{\label{fig6b} Stationary phase diagram of the process (\ref{rates3}) for $c_0 =2$, $c_1 =1$.
The phases $F(E)$, $F/C$ and $C/F$ are defined in Section 2, $F(E)$ and $F/C$ overlap (shaded region) and phase $F$ is empty.}
\end{figure}

As in the original model, the case $a=0$ leads to a continuous phase transition and this line of the phase diagram is identical to Figure \ref{fig1}. But for all $a\in (0,1)$, $\rho_c (a)>\rho_c (0)$, which is the value in (\ref{rhoc}) for the original model. So the phase region $F(E)$ in the phase diagram is larger than in the original model (see Figure \ref{fig6b}). To complete the phase diagram, we have to derive the analogue of the transition line $\rho_c +a$, which we call $\rho_{meta}$ in the following. This is defined by the emergence of a metastable condensed phase, characterized by a second local maximum of the rate function $I_\rho (\rho_{bg})$ in Theorem \ref{theo4}. The proof of this theorem makes use of the grand-canonical measures, and since these are now of different form, it does not apply directly. However, with some effort the proof can be written purely in terms of canonical measures (not shown here), and so the result (\ref{ratefct}) still applies, of course with $a$ replaced by $\rho\, a$. An analysis similar to Section 5 reveals that the rate function has an additional local minimum
\bea\label{irhoca}
I_\rho \big(\rho_c (0)\big)\quad\mbox{for }\rho >\rho_{meta} =\frac{\rho_c (0)}{1-a} \ .
\eea
So there exists a metastable condensed phase with background density $\rho_c (0)$, which is still the same as in the previous model, independent of $a$. This is to be expected, since the outflow of the condensate site has to match the background current. A simple heuristic argument along these lines provides a general framework to understand the transition, and is presented in detail in \cite{tobepublished}. Note that in comparison with (\ref{rhoca}),
\bea
\rho_{meta} (a)\leq \rho_c (a)\quad\mbox{for all }a\in [0,1)\ ,
\eea
with equality if and only if $a=0$. This follows immediately from the elementary inequality $x^{1-a} -1\leq (1-a)(x-1)$. So the phase regions $F(E)$ and $F/C$ as defined in Section 2 overlap (see shaded region in Figure \ref{fig6b}), and the region $F$ is empty. In contrast to our previous model, the equivalence of ensembles still holds in the presence of a metastable condensed phase.

\section{Discussion}

\subsection{Differences to previous results}

In the following we discuss differences in the condensation transition between zero-range processes with and without size-dependence in the jump rates. To simplify matters we concentrate on the rates (\ref{rates}) for $L$-dependent jump rates, but the features we discuss should hold in general.

\begin{itemize}

\item Without $L$-dependence the condensation transition in zero-range pro\-cesses is continuous, i.e. the background density $\rho_{bg}$ is a continuous function of the total particle density $\rho$. For model (\ref{rates}) this is only true if $a=0$, for $a>0$ the background density $\rho_{bg} =\rho_c <\rho_{trans}$ is smaller than the transition density and the transition is discontinuous.

\item If the jump rates do not depend on $L$, the equivalence of ensembles holds for all densities, and for $\rho\geq\rho_c$ the entropy density is linear in $\rho$ which is often characterized as partial equivalence of ensembles \cite{ellisetal00,touchetteetal04}. The reason is that the contribution of the condensate to the entropy density vanishes as $L\to\infty$. In model (\ref{rates}) this contribution does not vanish, cf. Theorem \ref{theo1}, and therefore we have only equivalence of ensembles for $\rho\leq\rho_c$ and nonequivalence for larger densities. As a consequence of this, the canonical entropy density is non-concave, whereas it is concave in case of no $L$-dependence.

\item Another striking feature of model (\ref{rates}) is that it exhibits ergodicity breaking, i.e. for $\rho >\rho_c +a$ there are two phases, fluid and condensed, with life-times exponential in $L$, one of which is metastable depending on the density. Without $L$-dependence in the jump rates this does not occur, and for all densities there is only one stable phase, either fluid for $\rho\leq\rho_c$ or condensed for $\rho >\rho_c$.

\end{itemize}

\noindent So far a discontinuous transition in a zero-range process has only been observed heuristically in a two-species system where the stationary state is not known \cite{godreche06}. The above features only concern the stationary measure, and for systems without $L$-depen\-dence they have been shown rigorously in a general context \cite{stefan2}. In the following we comment on further differences regarding equilibration and stationary dynamics, which have been studied only heuristically so far.

\begin{itemize}

\item If we prepare a system without $L$-dependence in a homogeneous distribution with density $\rho >\rho_c$ it exhibits coarsening \cite{stefan,godreche03}. Initially, clusters form all over the lattice, and as time progresses the larger cluster sites gain particles on the expense of the smaller cluster, leading to a self-similar time evolution. The driving force for this behaviour is the fact that there is no stable fluid phase with density $\rho >\rho_c$. This is not the case in model (\ref{rates}), which does not exhibit coarsening for that reason. Instead, it takes a time of order $e^{\xi^{fluid} L}$ before the condensate appears.

\item In a similar setting metastability has been reported 
as a precursor of the coarsening regime, i.e. before coarsening to a single
condensate sets in \cite{Kaup05}. Unlike in the present case, heuristic 
theoretical analysis supported by Monte-Carlo simulation shows that 
the life time of these metastable configurations does not grow exponentially 
with system size.

\item For systems without $L$-dependence, in the condensed phase the distribution of the homogeneous background has a sub-exponential tail \cite{stefan2}. In connection to this, the stationary time scale for movement of the condensate location 
(once a single condensate has build up)
is also sub-exponential in $L$, as was found heuristically in \cite{godrecheetal05} in case of a power law. For model (\ref{rates}) the background distribution is just $\nu_{c_1 ,\infty}^1 (k)\sim (c_1 /c_0 )^{-k}$ (see Theorem \ref{theo3}), which has an exponential tail. Therefore condensates can move only by dissolving completely and, after a time of order $e^{\xi^{fluid} L}$ in the coexisting fluid phase, forming on a different site. So the time scale for the stationary motion of a condensate is exponential in the system size.

\end{itemize}

\noindent The long time it takes to form a condensate in the present model
is observed in Monte Carlo simulations and is explained heuristically by a random walk picture in Section 5. The time it takes for the transition between the phases depends on the specific model as well as the definition of the phases. In any case its order is subexponential in the system size, and for the model (\ref{rates}) it is actually of order $L$. Moreover, if $\rho >\rho_c +na$ for $n\geq 1$ also more than one condensate is possible. But as can be seen in the proof of Theorem \ref{theo1}, the contribution to the partition function of such a configuration is negligible. Therefore one typically observes only one condensate, which is a common feature with the stationary behaviour of a system without $L$-dependence, although both cases have very different life times. In \cite{Schu07} a hydrodynamic
theory is developed for the time evolution under Eulerian scaling above the
condensation threshold. This leads to a generic picture for the evolution
of a space-dependent initial density profile with total supercritical density
in systems with rates that do not depend on $L$.
It would be interesting to study this problem in the present model.

Finally, we would also like to stress an intriguing difference to the usual theory of first order phase transitions in statistical mechanics. In systems with finite local state space or with bounded Hamiltonians, such as spin systems (Ising model) or exclusion models, the pressure $p$ is defined for all fugacities $\phi\geq 0$, and a first order phase transition is a result of the pressure being non-analytic (see e.g. \cite{ruelle69,varadhan88}). In the model we studied here, the pressure (\ref{pgcan}) is defined only for $\phi <c_1$, but is analytic on its domain. Therefore the phase transition is a result of this bounded domain in connection with the conservation of the particle number, and cannot be understood by studying the grand-canonical measures alone. This is in contrast to previously studied systems without $L$-dependent jump rates, where the presence of condensation can be characterized by a closed domain of the pressure $p$ (cf. \cite{stefan2}).

\subsection{Summary}

In this paper we presented a rigorous analysis of a discontinuous phase transition in a simple zero-range process with size-dependent jump rates. The model acts as a prototype for systems with that feature and the results are expected to be qualitatively similar for a large class of models. This will be discussed in detail in a forthcoming publication \cite{tobepublished}. Going
beyond earlier heuristic discussions of the phase transition
in terms of the order parameter 
\cite{lipowskietal02,coppexetal02,eggers99}, our analysis 
provides a detailed picture of the phase transition
in terms of the entire ensemble. In particular we note that
our approach is a pure equilibrium description, based as in
the work of \cite{lipowskietal02,coppexetal02} on the notion
of thermally activated processes. Hence 
there is no need for an appeal \cite{eggers99} to a non-equilibrium 
dissipative structure, maintained 
by a flux of entropy, for understanding the nature of the
condensation transition in the granular shaking experiment.

Since our results only concern the stationary distribution they do not depend on the geometry or dimension of the lattice. The model shows the same features observed in granular clustering, namely metastability and a first order transition. The only difference is that there is no region in the phase diagram where the fluid phase becomes unstable. This is due to the simple choice of rates in this first analysis, and the issue will be addressed in \cite{tobepublished}.

\section*{Acknowledgments}
Both authors would like to thank Pablo Ferrari for inspiring discussions and useful comments about essential parts of the manuscript, and for an invitation to NUMEC at the University of S\~ao Paulo, which was supported by FAPESP. The authors are also grateful for the hospitality of the Isaac Newton Institute in Cambridge, where this work was initiated during the
programme ``Principles of the Dynamics of Non-Equilibrium 
Systems''.

\appendix

\section*{Appendix}
\setcounter{section}{1}
\subsection{Proof of Proposition \ref{prop1}}
For $\rho <\rho_c$ (\ref{nlln}) follows by standard results and for $\rho >\rho_c$ we have as $L\to\infty$
\bea
\nu^1_{\phi_R (\rho ),R} (\eta_x^L >R) \simeq\sqrt{\frac{\rho -\rho_c }{z_\infty (c_1 )}}\Big(\frac{c_1}{c_0}\Big)^{R/2} \ .
\eea
Therefore if we define the truncated occupation numbers $\hat\eta_x^L =\eta_x^L \wedge R$, we have
\bea
\lefteqn{\nu^L_{\phi_R (\rho ),R} \Big(\frac1L\sum_{x\in\Lambda_L} (\eta_x^L -\hat\eta_x^L )\neq 0\Big) =\nu^L_{\phi_R (\rho ),R} \big(\mbox{at least one }\eta_x^L >R\big) =}\nonumber\\
& &\quad =1-\Big( 1-\nu^1_{\phi_R (\rho ),R} (\eta_x >R)\Big)^L \leq C L\Big(\frac{c_1}{c_0}\Big)^{R/2} \ .
\eea
With $R\gg\log L$ this bound is summable %, i.e.
%\bea
%\sum_{L=1}^\infty \nu^L_{\phi_R (\rho ),R} \Big(\sum_{x\in\Lambda_L} (\eta_x^L -\hat\eta_x^L )\neq 0\Big) <\infty
%\eea
and the Borel-Cantelli Lemma implies that
\bea
\frac1L\sum_{x\in\Lambda_L} \eta_x^L - \frac1L\sum_{x\in\Lambda_L} \hat\eta_x^L \to 0\quad a.s.\quad\mbox{as }L\to\infty .
\eea
Moreover $\langle\hat\eta_x \rangle =\rho_c +O\big( (\frac{c_1}{c_0})^{R/2}\big)$ and $Var (\hat\eta_x )\leq\frac{c_0 c_1}{(c_0 -c_1 )^2} +O\big( R(\frac{c_1}{c_0})^{R/2}\big)$ and therefore by the usual strong law we have $\sum_{x\in\Lambda_L} \hat\eta_x^L \to\rho_c \quad a.s.$\ .\\
Taken together, this implies (\ref{nlln}) for $\rho >\rho_c$, and $\rho =\rho_c$ works analogously with the power $R/2$ replaced by $R/4$.
%For $\rho >\rho_c$ (\ref{nlln}) follows by convergence of the probability generating function
%\bea
%\mathit{pgf}_{\sum\eta_x /L} (s)&=&\big\langle s^{\eta_x /L} \big\rangle_{\nu^1_{\phi_R ,R}}^L =\bigg( \frac{z_R (\phi_R s^{1/L})}{z_R (\phi_R )}\bigg)^L \simeq\nonumber\\
%&\simeq&\bigg( \frac{c_0 -\phi_R}{c_0 -\phi_R s^{1/L}} \bigg)^L \bigg( \frac{R+1+s^a (\rho -\rho_c )}{R+1+\rho -\rho_c }\bigg)^L \nonumber\\
%&\to & s^{\rho_c} \exp\Big( (s^a -1)\frac{\rho -\rho_c}{a} \Big)\quad\mbox{as }L\to\infty\ .
%\eea

\subsection{Proof of Lemma \ref{lemma1}}
Each configuration in $X_{L,N} \setminus X^0_{L,N}$ has at least one site with more than $R$ particles and we denote the number of such sites by
\bea
E(\feta ):=\sum_{x\in\Lambda_L} \1_{\eta_x >R} (\feta )\ .
\eea
Note that for $\feta\in X_{L,N} \setminus X^0_{L,N}$ we have
\bea
1\leq E(\feta )\leq M=\lceil N/R\rceil\ ,
\eea
where $M$ is as defined in (\ref{decompose}). For each configuration we define
%$\feta\mapsto \bar\feta\mbox{where} \bar\eta_x =\eta_x\wedge R
\bea
S(\feta ):=\big(\eta_x \wedge R\,\big|\, x{\in}\Lambda_L \big) \cup \big(\eta_x {-}R\,\big|\, x{\in}\Lambda_L ,\,\eta_x {>}R\big)\in X_{L{+}E(\feta ),N}\ .
\eea
If $E\big( S(\feta )\big) >0$, we have to repeat this mapping at most $M$ times such that
\bea
\bar\feta :=S^M (\feta )\in X_{L+l(\feta ),N}^0 \ ,
\eea 
where $l(\feta )\leq M$ denotes the total number of extra coordinates. For $l(\feta )< M$ we can identify $\bar\feta$ by a configuration in $X_{L+M,N}^0$, by setting all remaining coordinates equal to zero. By this construction it is clear that for each $\feta\in X_{L,N} \setminus X^0_{L,N}$ there exists a unique $\bar\feta \in X_{L+M,N}^0$, i.e.
\bea
\big| X_{L,N} \setminus X^0_{L,N} \big| =\big| X_{L,N} \big| -\big| X^0_{L,N} \big|\leq \big| X_{L+M,N}^0 \big|\ .
\eea
Further, each $\bar\feta$ has the special property that at least $l(\feta )$ sites contain exactly $R$ particles, and there are only $L$ sites whose occupation number can be less or equal than that. Therefore we can improve the above estimate as
\bea
\big| X_{L,N} \big| -\big| X^0_{L,N} \big|\leq {L+M\choose M}\,\big| X_{L,N-R}^0 \big|\leq {L+M\choose M}\,\frac{\big| X_{L,N}^0 \big|}{(L-M)^R}\ ,
\eea
where the combinatorial factor counts the number of positions of sites with $R$ particles. We also used the fact that for all $k=1,\ldots ,R$
\bea
\big| X_{L,N-k+1}^0 \big|\geq (L-M) \big| X_{L,N-k}^0 \big|\ ,
\eea
since there are at least $L-M$ positions to put an additional particle without violating the constraint $\eta_x \leq R$ for all $x$. Together with the obvious fact that $\big| X^0_{L,N} \big|\leq \big| X_{L,N} \big|$, this proves the first statement of the lemma, i.e.
\bea\label{stirli}
\frac{1}{1+{L+M\choose M}\big/(L-M)^R}\, \big| X_{L,N} \big|\leq\big| X^0_{L,N} \big|\leq \big| X_{L,N} \big|\ .
\eea
With Stirling's formula we get
\bea
\lim_{L\to\infty} \frac1L\log {L{+}M\choose L}=\lim_{L\to\infty}\Big(\big( 1{+}\tfrac{M}{L}\big)\log \big( 1{+}\tfrac{M}{L}\big) {-}\tfrac{M}{L}\log\tfrac{M}{L}\Big) =0\ .
\eea
since $M/L\to 0$ as $L\to\infty$. Therefore
\bea
\frac1L\log\frac{1}{1+{L+M\choose M}\big/(L-M)^R}\to 0
\eea
and (\ref{stirli}) certainly includes the second statement of the lemma. More detailed, we get to leading order as $L\to\infty$, $N/L\to\rho$,
\bea
\frac{{L+M\choose M}}{(L-M)^R}=\Big(\frac{R+\rho}{\rho}\Big)^{\rho L/R+1/2}\Big( 1+\frac{\rho}{R}\Big)^L \frac{L^{-R-1/2}}{\sqrt{2\pi}}\big( 1+o(1)\big)\ .
\eea
This vanishes for all $\rho\geq 0$ if
\bea
R\log L-\frac{\rho L}{R}\log R\gg \log R\vee \frac{\rho L}{R}\quad\mbox{as }L\to\infty\ ,
\eea
which is certainly the case for $R\gg\sqrt{L}$.  
 \hfill $\Box$


\begin{thebibliography}{10}

\bibitem{andjel82}
Andjel, E.: Invariant measures for the zero range process. Ann.\ Probab.\ \textbf{10}(3), 525--547 (1982)

\bibitem{angeletal07}
Angel, A.G., Evans, M.R., Levine, E., Mukamel, D.: Criticality and Condensation in a Non-Conserving Zero Range Process. J.\ Stat.\ Mech.\ P08017 (2007)

\bibitem{bovier}
Bovier, A.: Metastability: a potential theoretic approach. In: Proceedings of the ICM 2006, Madrid, pp.\ 499--518, European Mathematical Society (2006)

\bibitem{coppexetal02}
Coppex, F., Droz, M., Lipowski, A.: Dynamics of the breakdown of granular clusters. Phys.\ Rev.\ E \textbf{66}, 011305 (2002)

\bibitem{csiszar75}
Csisz\'ar, I.: I-divergence geometry of probability distributions and minimization problems. Ann.\ Probab.\ \textbf{3}(1), 146-–158 (1975)

\bibitem{csiszar84}
Csisz\'ar, I.: Sanov property, generalized i-projection and a conditional limit theorem. Ann.\ Probab.\ \textbf{12}, 768--793 (1984)

\bibitem{eggers99}
Eggers, J.: Sand as Maxwell's Demon. Phys.\ Rev.\ Lett.\ \textbf{83}(25), 5322-5325 (1999)

\bibitem{ellisetal00}
Ellis, R.S., Haven, K., Turkington, B.: Large deviation principles and complete equivalence and nonequivalence results for pure and mixed ensembles. J.\ Stat.\ Phys.\ \textbf{101}, 999--1064 (2000)

\bibitem{evans00}
Evans, M.R.: Phase transitions in one-dimensional nonequilibrium systems. Braz.\ J.\ Phys.\ \textbf{30}(1), 42--57 (2000)

\bibitem{evansetal05}
Evans, M.R., Hanney, T.: Nonequilibrium statistical mechanics of the zero-range process and related models. J.\ Phys.\ A: Math.\ Gen.\ \textbf{38}, R195--R239 (2005)

%\bibitem{evansetal05b}
%Evans, M.R., Majumdar, S.N., Zia, R.K.P.: Canonical analysis of condensation in factorised steady state. J.\ Stat.\ Phys.\ \textbf{123}, 357--390 (2006)

\bibitem{evansetal06}
Evans, M.R., Hanney, T., Majumdar, S.N.: Interaction driven real-space condensation. Phys.\ Rev.\ Lett.\ \textbf{97}, 010602 (2006)

\bibitem{ferrarietal07}
Ferrari, P.A., Landim, C., Sisko, V.V.: Condensation for a fixed number of independent random variables. J.\ Stat.\ Phys.\ \textbf{128}, 1153-1158 (2007)

\bibitem{godreche03}
Godr\`eche, C.: Dynamics of condensation in zero-range processes. J.\ Phys.\ A: Math.\ Gen.\ \textbf{36}, 6313-6328 (2003)

\bibitem{godrecheetal05}
Godr\`eche, C., Luck, J.M.: Dynamics of the condensate in zero-range processes. J.\ Phys.\ A: Math.\ Gen.\ \textbf{38}, 7215-7237 (2005)

\bibitem{godreche06}
Godr\`eche, C.: Nonequilibrium phase transition in a non integrable zero-range process. J.\ Phys.\ A: Math.\ Gen.\ \textbf{39}, 9055--9069 (2006)

\bibitem{stefan2}
Grosskinsky, S.:  Equivalence of ensembles for two-component zero-range invariant measures. accepted in Stoch.~Proc.~Appl. (DOI: 10.1016/j.spa.2007.09.006)

\bibitem{stefan}
Grosskinsky, S., Sch\"utz, G.M., Spohn, H.: Condensation in the zero range process: stationary and dynamical properties. J.\ Stat.\ Phys.\ \textbf{113}(3/4), 389--410 (2003)

\bibitem{tobepublished}
Grosskinsky, S., Sch\"utz, G.M.: in preparation

%\bibitem{jeonetal00}
%Jeon, I., March, P., Pittel, B.: Size of the largest cluster under zero-range invariant measures. Ann.\ Probab.\ \textbf{28}(3), 1162--1194 (2000)

\bibitem{kafrietal02}
Kafri, Y., Levine, E., Mukamel, D., Sch\"utz, G.M., T\"or\"ok, J.: Criterion for phase separation in one-dimensional driven systems. Phys.\ Rev.\ Lett. \textbf{89}(3), 035702 (2002)

\bibitem{Kaup05} Kaupuzs, J., Mahnke, R., Harris, R.J.: Zero-range model of traffic flow. 
Phys.\ Rev.\ E \textbf{72}(5), 056125 (2005)

%\bibitem{kipnislandim}
%Kipnis, C., Landim, C.: Scaling Limits of Interacting Particle Systems. Volume 320 of Grundlehren der mathematischen Wissenschaften, Springer Verlag, Berlin (1999)

\bibitem{levineetal05}
Levine, E., Mukamel, D., Sch\"utz, G.M.: Zero range process with open boudaries. J.\ Stat.\ Phys.\ \textbf{120}(5/6), 759-778 (2002)

\bibitem{liggett}
Liggett, T.M.: Interacting Particle Systems. Springer, Berlin (2004)

\bibitem{lipowskietal02}
Lipowski, A., Droz, M.: Urn model of separation of sand. Phys.\ Rev.\ E \textbf{65}, 031307 (2002)

\bibitem{loulakisetal07}
Loulakis, M., Armend\'{a}riz, I.: Thermodynamic Limit for the Invariant Measures in Supercritical Zero Range Processes. arXiv:0801.2511 (2008)

\bibitem{ruelle69}
Ruelle, D.: Statistical mechanics: rigorous results. W.A.\ Benjamin, New
  York-Amsterdam (1969)

\bibitem{schlichtingetal96} 
Schlichting, H.J., Nordmeier, V.: Strukturen im Sand. Kollektives Verhalten und Selbstorganisation bei Granulaten. Math.\ naturw.\ Unterricht \textbf{49}6, 323-332 (1996) (in German)

\bibitem{Schu07}
Sch\"utz, G.M., Harris, R.J.: Hydrodynamics of the zero-range process in the condensation regime. J.\ Stat.\ Phys.\ \textbf{127}(2), 419-430 (2007)

\bibitem{schwarzkopfetal08}
Schwarzkopf, Y., Evans, M.R., Mukamel, D.: Zero-Range Processes with Multiple Condensates: Statics and Dynamics. arXiv:0801.4501 (2008)

\bibitem{spitzer70}
Spitzer, F.: Interaction of markov processes. Adv.\ Math. \textbf{5}, 246--290 (1970)

\bibitem{toeroek05}
T\"or\"ok, J.: Analytic study of clustering in shaken granular material using zero-range processes. Physica A \textbf{355}, 374--382 (2005)

\bibitem{touchetteetal04}
Touchette, H., Ellis, R.S., Turkington, B.: An introduction to the thermodynamic and macrostate levels of nonequivalent ensembles. Physica A \textbf{340}, 138--146 (2004)

\bibitem{meeretal02}
van der Meer, D., van der Weele, J.P., Lohse, D.: Sudden collapse of a granular cluster. Phys.\ Rev.\ Lett.\ \textbf{88}, 174302 (2002)

\bibitem{meeretal07}
van der Meer, D., van der Weele, K., Reimann, P., Lohse, D.: Compartmentalized granular gases: flux model results. J. Stat. Mech.: Theor. Exp. P07021 (2007)

\bibitem{weeleetal01}
van der Weele, J.P., van der Meer, D., Versluis, M., Lohse, D.: Hysteretic clustering in granular gas. Europh.\ Lett.\ \textbf{53}(3), 328-334 (2001)

\bibitem{varadhan88}
Varadhan, S.R.S: Large deviations and applications. Ecole d'Et\'e de Probabilit\'es de Saint-Flour XV-XVII. Lecture Notes in Math., Springer, Berlin (1988)

\end{thebibliography}
\end{document}